%% file: excl_paper.tex
\pdfoutput=1
\newcommand*{\ATLASLATEXPATH}{./}
\newcommand*{\emu}{\ensuremath{e^{\pm}\mu^{\mp}}}
\newcommand*{\WW}{\ensuremath{W^+W^-}}
\newcommand*{\yyWW}{\ensuremath{\gamma\gamma \to W^+W^-}}
\newcommand*{\yyll}{\ensuremath{\gamma\gamma \to \ell^+\ell^-}}
\newcommand*{\yymumu}{\ensuremath{\gamma\gamma \to \mu^+\mu^-}}
\newcommand*{\yyee}{\ensuremath{\gamma\gamma \to e^+e^-}}
\newcommand*{\yytautau}{\ensuremath{\gamma\gamma \to \tau^+\tau^-}}
\newcommand*{\gamgamWWll}{\ensuremath{\gamma\gamma \to W^+W^- \to \ell^{+} \nu \ell'^{-} \bar{\nu}}}
\newcommand*{\ggHWWll}{\ensuremath{gg\ra H \ra W^+W^- \ra \ell^{+} \nu \ell'^{-} \bar{\nu}}}
\newcommand*{\DrellYantomumu}{\ensuremath{Z/\gamma^* \to \mu^+\mu^-}}
\newcommand*{\DrellYantotautau}{\ensuremath{Z/\gamma^* \to \tau^+\tau^-}}
\newcommand*{\pTll}{\ensuremath{p_{\textrm{T}}^{\ell\ell}}}

\newcommand*{\pTemu}{\ensuremath{p_{\textrm{T}}^{e\mu}}}
\newcommand*{\pTmumu}{\ensuremath{p_{\textrm{T}}^{\mu\mu}}}

\newcommand*{\memu}{\ensuremath{m_{e\mu}}}
\newcommand*{\mmumu}{\ensuremath{m_{\mu\mu}}}
\newcommand*{\mT}{\ensuremath{m_{\textrm{T}}}}

\newcommand*{\dFem}{\ensuremath{\Delta\phi_{e\mu}}}

\newcommand*{\DZ}{\ensuremath{\Delta z_0^{\mathrm{iso}}}}
\newcommand*{\fEL}{\ensuremath{f_{\mathrm{EL}}}}
\newcommand*{\fgam}{\ensuremath{f_{\gamma}}}
\newcommand*{\fP}{\ensuremath{f^{\mathrm{P}}}}
\newcommand*{\fAP}{\ensuremath{f^{\mathrm{AP}}}}
\newcommand*{\fAH}{\ensuremath{f^{\mathrm{AH}}}}
\newcommand*{\LambdaCutoff}{\ensuremath{\Lambda_{\textrm{cutoff}}}}
\newcommand*{\BR}{\textrm{BR}}
\newcommand*{\mm}{\ensuremath{\mathrm{mm}}}
\newcommand*{\ggww}{\textsc{gg2ww}\xspace}
\newcommand*{\PowhegPythiaEight}{\textsc{Powheg+Pythia8}\xspace}
\newcommand*{\AlpgenPythiaSix}{\textsc{Alpgen+Pythia6}\xspace}
\newcommand*{\AlpgenHerwig}{\textsc{Alpgen+Herwig}\xspace}
\newcommand*{\PYTHIAsix}{\textsc{Pythia6}\xspace}

\documentclass[USenglish,texlive=2011,PAPER,txfonts=true,cernpreprint]{\ATLASLATEXPATH atlasdoc}

\usepackage[subfigure, block=none,default]{\ATLASLATEXPATH atlaspackage}
\usepackage{multirow,bigdelim}
\usepackage{\ATLASLATEXPATH atlasbiblatex}

\usepackage{\ATLASLATEXPATH atlascontribute}

\usepackage{\ATLASLATEXPATH atlasphysics}
\usepackage{\ATLASLATEXPATH atlasparticle}

\graphicspath{{./}{./}}

 \AtlasJournalRef{\PRD 94 (2016) 032011}
 \AtlasDOI{10.1103/PhysRevD.94.032011}

\AtlasTitle{Measurement of exclusive $\gamma\gamma\rightarrow W^+W^-$ production and search for exclusive 
Higgs boson production in $pp$ collisions at $\sqs = 8~\TeV$ using the ATLAS detector}
\author{The ATLAS Collaboration}
\AtlasRefCode{STDM-2015-10}
 \AtlasNote{ATL-COM-PHYS-2016-082}
\PreprintIdNumber{CERN-EP-2016-123}
 \AtlasJournal{Phys.\ Rev.\ D.}
\AtlasAbstract{%
 Searches for exclusively produced $W$ boson pairs in the process $pp(\gamma\gamma) \rightarrow pW^+W^-p$  
and exclusively produced Higgs boson in the process $pp(gg) \rightarrow pHp$ have been performed using $e^{\pm}\mu^{\mp}$ final states.  
These measurements use 20.2~\ifb\ of $pp$ collisions collected by the 
ATLAS experiment at a center-of-mass energy $\sqrt{s}=8$~TeV 
at the LHC. Exclusive production of $W^+W^-$ consistent with the
Standard Model  
prediction is found with 3.0$\sigma$ significance. The exclusive $W^+W^-$ production cross-section 
is determined to be
$\sigma (\gamma\gamma\rightarrow W^{+}W^{-}\rightarrow e^{\pm}\mu^{\mp} X)
= 6.9 \pm 2.2 (\text{stat.}) \pm 1.4 (\text{sys.})$~fb, 
in agreement with the Standard Model prediction. Limits on anomalous quartic gauge couplings 
are set at 95\% confidence-level as $-1.7 \times 10^{-6} < a_0^W/\Lambda^2 < 1.7 \times 10^{-6}$ GeV$^{-2}$
and $-6.4 \times 10^{-6} < a_C^W/\Lambda^2 < 6.3 \times 10^{-6}$
GeV$^{-2}$.  
A 95\% confidence-level upper limit on the total production cross-section for exclusive Higgs boson is set to 1.2~pb.
}
 \AtlasCoverSupportingNote{Internal Supporting Document}{https://cds.cern.ch/record/2007521}
 \AtlasCoverCommentsDeadline{19 May 2016}
 \AtlasCoverAnalysisTeam{Chav Chhiv Chau, Last Feremenga, Lawrence Nodulman, William Trischuk, Jaehoon Yu, Sergei Chekanov, Justin Griffiths, Robert Blair}
 \AtlasCoverEdBoardMember{Laurent Olivier Schoeffel~(chair)}
 \AtlasCoverEdBoardMember{Brian Edward Lindquist}
 \AtlasCoverEdBoardMember{Gideon Bella}
 \AtlasCoverEgroupEditors{atlas-stdm-2015-10-editors@cern.ch}
 \AtlasCoverEgroupEdBoard{atlas-stdm-2015-10-editorial-board@cern.ch}

\begin{document}

\maketitle

\tableofcontents


\section{Introduction}
\label{sec:intro}
\par In the Standard Model (SM) of particle physics, the interactions between electroweak gauge bosons
are described by the non-Abelian $SU(2) \times U(1)$ structure of the electroweak sector.
Measurement of the strengths of the trilinear ($VVV$, where $V$ = $\gamma$, $W$, or $Z$) and 
quartic ($VVVV$) gauge couplings represent an important test of the SM, as deviations 
from SM predictions would indicate new physics.
The discovery of a Higgs boson~\cite{Aad2012tfa,Chatrchyan:2012xdj} at the Large Hadron Collider (LHC)
has taken a major step toward confirming the mechanism of electroweak symmetry breaking. 
Anomalous quartic gauge couplings (aQGCs) provide a window to further probe 
possible new physics extensions of electroweak theory.  Exclusive  
production of $W$ boson pairs, $pp(\gamma\gamma) \to p \WW p$, provides an opportunity to study 
$\yyWW$ aQGC couplings~\cite{Belanger,Chapon}.
\par
\begin{figure}
 \subfigure[Elastic production]{
   \includegraphics[width=0.30\textwidth]{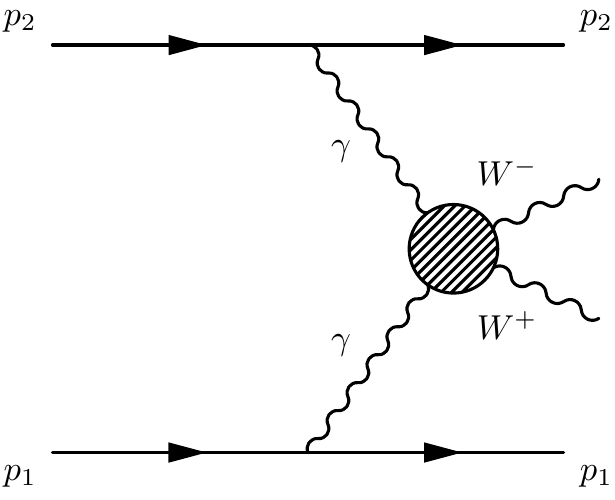}
   \label{fig:yyWW-elastic}
 }\hfill%
 \subfigure[{Single-dissociation}]{
   \includegraphics[width=0.30\textwidth]{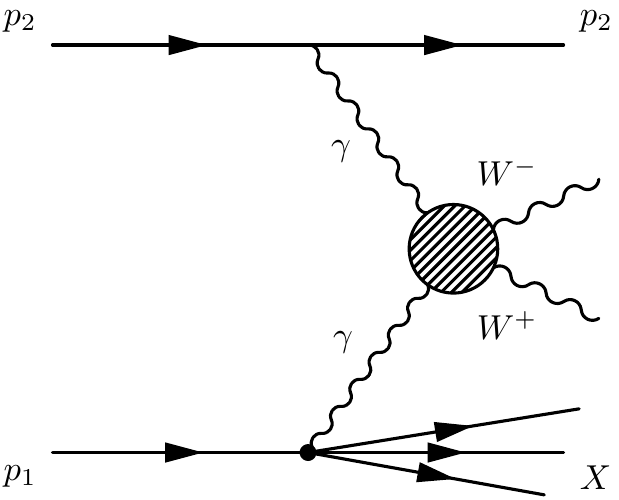}
   \label{fig:yyWW-sd}
 }\hfill%
 \subfigure[{Double-dissociation}]{
   \includegraphics[width=0.30\textwidth]{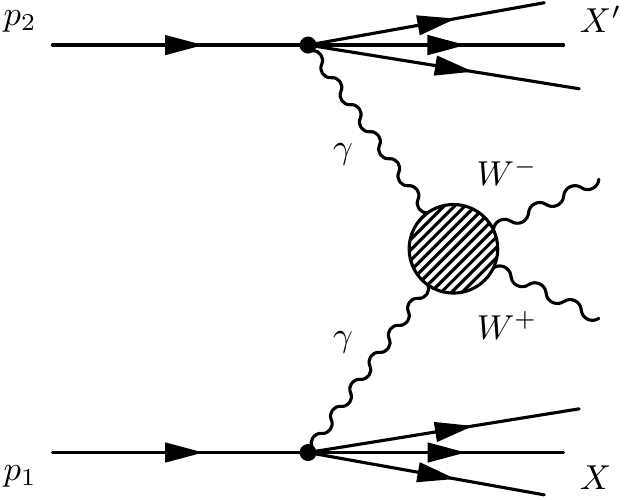}
   \label{fig:yyWW-dd}
 }
 \caption{Diagrams for the exclusive $\yyWW$ production
          representing the (a) elastic process, 
                           (b) single-dissociation where one initial proton dissociates (SD)
                       and (c) double-dissociation where both protons fragment (DD).
          The symbols $X$ and $X'$ denote any additional final state created.}
 \label{fig:exclWW-diagram}
\end{figure}

 In $pp$ collisions, exclusive $\WW$ events are produced when each proton 
emits a photon and the two photons annihilate, either via $t$- and $u$-channel 
$W$-exchange diagrams involving trilinear gauge couplings
or via quartic gauge coupling diagram, to create a $\WW$ pair.
Figure~\ref{fig:exclWW-diagram} shows the exclusive production of a  
$\WW$ pair, where the blobs represent the $t$-channel, $u$-channel, and quartic diagrams.
After the collisions, either both protons remain intact as shown in 
Fig.~\ref{fig:yyWW-elastic} (referred to as \emph{elastic} hereafter),
only one proton remains intact as in Fig.~\ref{fig:yyWW-sd} (\emph{single-dissociation}, SD), 
or both protons dissociate as in Fig.~\ref{fig:yyWW-dd} (\emph{double-dissociation}, DD). 
In all three cases the trajectories of the protons or their
remnants deviate only slightly from 
their initial directions so that they never enter the acceptance of the ATLAS detector. 
On the other hand, inclusive processes are produced with accompanying activity such as
initial- and final-state radiation and additional scattering in the same $pp$ collision.
The accompanying activity is collectively called the underlying event and emits particles 
into the acceptance of the ATLAS detector. 
\par Photon scattering in hadron colliders can be described in quantum electrodynamics (QED) by
the equivalent-photon approximation (EPA)~\cite{EPA0,EPA}. In this framework the 
exclusive $\WW$ cross-section can be written as  
\begin{equation}
  \sigma_{pp(\gamma\gamma) \to pp\WW}^{\textrm{EPA}} = \iint f(x_1) f(x_2) \sigma_{\gamma\gamma\to \WW} (m_{\gamma\gamma}^2) dx_1 dx_2,
  \label{eqn:EPAxsection}
\end{equation}
where $f(x_i)$, for $i\in\{1,2\}$, is the number of equivalent photons carrying a fraction of the proton's 
energy, $x_i$, that are emitted, while
$m_{\gamma\gamma}$ is the two-photon center-of-mass energy.
This approach has been used to describe similar exclusive processes in 
the CDF~\cite{CDFgg},
STAR~\cite{STARgg}, and CMS~\cite{CMSmumu,CMSee} experiments.
\par Exclusive $W^+W^-$ pair production is particularly sensitive to new 
physics that may be described by anomalous quartic gauge coupling (aQGC) 
of the form $WW\gamma\gamma$~\cite{KepkaRoyon,Chapon}. 
The dimension-6 operators in Ref.~\cite{Belanger} are the lowest-dimension operators that give
rise to anomalous $WW\gamma\gamma$ couplings, $a_0^W/\Lambda^2$ and
$a_C^W/\Lambda^2$ where $\Lambda$ is the scale of new physics.
A procedure adopted by previous measurements~\cite{OPALaQGC, D0WWyy, CMSyyWW7and8} 
uses a dipole form factor to preserve unitarity at high  $m_{\gamma\gamma}$. 
The couplings $a_0^W / \Lambda^2$ and $a_C^W / \Lambda^2$ then become:
\begin{equation}
 a_{0,C}^W / \Lambda^2 \to \frac{a_{0,C}^W}{\Lambda^2} \frac{1}{\left(1 + \frac{m_{\gamma\gamma}^2}{\Lambda_{\mathrm{cutoff}}^2}\right)^2} 
 \label{eqn:dipoleff}
\end{equation}
where $\Lambda_{\mathrm{cutoff}}$ defines the scale of possible new
physics, and the term containing it ensures that unitarity is preserved.
\par Anomalous triple gauge
couplings (aTGCs) could also produce similar effects but the sensitivity of
this study to aTGCs is
not competitive compared with other processes~\cite{Chapon}, so these are taken to be zero. 
\par More recent
parameterizations of aQGCs are of dimension 8. 
The parameterization of the dimension-8  
couplings, $f_{M,0,1,2,3}/\Lambda^{4}$, in Ref.~\cite{SnowmassPaper} are linearly related to the 
$a_{0,C}^W / \Lambda^2$ as follows:
\begin{equation}
 \frac{f_{M,0}}{\Lambda^{4}} = \frac{a_0^W}{\Lambda^{2}} \frac{1}{g^2 \nu^2}, \qquad 
     \frac{f_{M,1}}{\Lambda^{4}} = -\frac{a_C^W}{\Lambda^{2}} \frac{1}{g^2 \nu^2}, 
  \label{eqn:relationD8toD6}
\end{equation}
where $g = e/\sin(\theta_{W})$ and $\nu$ is the Higgs boson vacuum expectation value. Also, with 
this parameterization, $f_{M,2} = 2 \times f_{M,0}$ and  $f_{M,3} = 2 \times f_{M,1}$.
\par In addition to the discovery of the Higgs boson, several of its properties -- such as mass,
 coupling strengths to various final-state particles, and branching ratios of its decay --  have 
been determined~\cite{Aad2012tfa, Aad:2015ona} using Higgs boson candidates from inclusive production. 
Higgs boson candidates from the exclusive production
$(pp\rightarrow pggp\rightarrow pHp)$ would have lower systematic uncertainties due to their cleaner 
production environment~\cite{Heinemeyer:2007tu, Cox:2007sw, Tasevsky:2014cpa,Cox:2005if}. 
Since measurements using these Higgs boson candidates would have better precision, they could be used 
to improve knowledge of the Higgs boson sector.  
It is therefore interesting to determine the cross-section for exclusive Higgs boson production and examine 
the feasibility of using exclusive Higgs boson candidates for Higgs
boson property measurements. 
This interest is reflected in the inclusion of the exclusive Higgs
boson process studies as 
part of the physics program of forward proton-tagging
detectors~\cite{Royon:2014kta, AFPTDR, TOTEMTDR} that extend the ATLAS and CMS 
coverage for LHC runs at 13~\TeV.
\par
\begin{figure}[!h]
\centering
\includegraphics[width=0.4\linewidth]{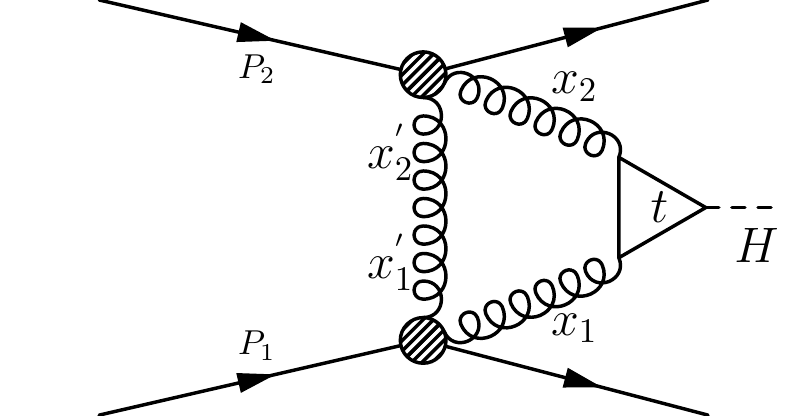}
\caption{The lowest-order Feynman diagram for the exclusive Higgs
boson production.
The variables $x_1$ and $x_2$ are the fractions of the momenta carried by the gluons that contribute 
to the production of the Higgs boson, with respect to the momenta of the 
protons $P_1$ and $P_2$. The variables $x_1^{'}$, and $x_2^{'}$, on the other hand, are the fractions 
of the momentum carried by the exchanged third gluon with respect to the momenta of the
protons $P_1$ and $P_2$.}
\label{fig:exclH}
\end{figure}

 Unlike exclusive $\WW$ production, exclusive Higgs boson production proceeds through 
a quantum chromodynamics (QCD) process involving at least three gluons as shown
in Fig.~\ref{fig:exclH}.
Two gluons from the colliding protons interact through a top-quark loop to 
produce a Higgs boson while additional gluon exchange between the colliding 
protons keeps the protons color-neutral and allows
the protons to remain intact 
after the collision. The proton 
trajectories deviate slightly after the collision. One $W$ boson from
Higgs boson decays must be off shell so the event selection for that study
needs to be different than the exclusive $\WW$ event selection, and the samples are largely orthogonal.  
\par The exclusive Higgs boson production cross-section can be written as~\cite{Khoze} 
\begin{equation}
\sigma_{pp(gg)\ra ppH} \propto \hat{\sigma}(gg\ra H)\left ( \int \frac{dQ^2_t}{Q^4_t}f_g(x_1, x_1^{'},Q^2_t)f_g(x_2, x_2^{'},Q^2_t)\right )^2 
\end{equation}
where $\hat{\sigma}(gg\ra H)$ is the cross-section
for the gluon fusion process that produces the Higgs boson. 
The functions $f_g$~\cite{Martin} are the generalized gluon densities
for the finite proton size, 
that take into account the impact parameter. 
The variables $x_1$ and $x_2$ are the fractions of the momenta carried by the gluons that contribute 
to the production of the Higgs boson, with respect to the momenta of the 
protons $P_1$ and $P_2$. The variables $x_1^{'}$ and $x_2^{'}$ are the fractions 
of the momentum carried by the exchanged third gluon with respect to the momenta of the
protons $P_1$ and $P_2$ as shown in Fig.~\ref{fig:exclH}.
These gluon densities are integrated over the exchanged (third) gluon 
transverse momentum $Q_t$. This formalism, used in several theoretical 
calculations, predicts cross-sections that vary by over an order of magnitude~\cite{Khoze,Cudell}.
This wide disparity in predictions is an additional motivation for
this measurement.
While either proton could dissociate, the predictions presented here are for elastic production only and 
could underestimate the cross-section by an order of magnitude~\cite{Khoze}.
\par This paper describes searches for exclusive $\WW$ and $H\to\WW$ production 
using  $e^{\pm}\mu^{\mp}$ final states.
Events where a $W$ boson decays to a $\tau$ lepton that subsequently decays to an electron or muon 
are also included. This final state is denoted $e\mu X$, where $X$ 
represents the neutrinos. 
Section~\ref{sec:detector} describes the experimental setup. Section~\ref{sec:dataMCsamples} describes the 
dataset and simulation tools used to model signal and background processes.
Initial selection of electron, muon, jet and track candidates is discussed in Section~\ref{sec:physicsObjects}.  
Section~\ref{sec:exclSelection} introduces a new approach to separate exclusive from 
inclusive production processes. 
Section~\ref{sec:eventSelection} describes the event selections including
signal regions                  
for both the exclusive $\WW$ and Higgs boson processes.
Section~\ref{sec:validationDilepton} outlines studies of the 
exclusive event selection and underlying-event models using samples of
same-flavor opposite-sign lepton pairs in
$p\gamma\gamma p \to p\ell^+\ell^- p$ candidates ($\ell = \mu$ or $e$)
to validate modeling and selection criteria. 
In Section~\ref{sec:controlRegions},
data control regions designed to test and correct physics and
detector modeling are described. Systematic 
uncertainties are summarized in Section~\ref{sec:systs} and the results of the study 
are described in detail in Section~\ref{sec:results}. Section~\ref{sec:conclusion} summarizes the findings.

\section{The ATLAS detector}
\label{sec:detector}
ATLAS~\cite{PERF-2007-01} is a multipurpose cylindrical detector\footnote{The ATLAS experiment uses a right-handed coordinate system with its origin at the nominal interaction point (IP) in the
center of the detector and the $z$-axis along the beam direction. The $x$-axis points from the IP to the center of the LHC ring
and the $y$-axis points upward. Cylindrical coordinates ($r,\phi$) are used in the transverse ($x,y$) plane,
$\phi$ being the azimuthal angle around the $z$-axis. The pseudorapidity is defined in terms of the polar angle
$\theta$ as $\eta=-\ln\tan(\theta/2)$. The angular distance $\Delta R$ in the $\eta$-$\phi$ space is defined as 
$\Delta R = \sqrt{(\Delta\eta)^2 + (\Delta\phi)^2}$. } 
 that consists of an inner detector surrounded by a superconducting solenoid, a calorimeter 
system, and a muon spectrometer that includes superconducting toroidal magnets. The inner detector system consists of three 
subsystems: a pixel detector, a silicon microstrip detector, and a transition radiation 
tracker. Immersed in a 2~T magnetic field provided by the superconducting solenoid, these three subsystems enable the 
inner detector to accurately reconstruct the trajectories of charged particles in a pseudorapidity range 
$\abseta<2.5$ and measure their momenta and charges. The inner detector is surrounded by high-granularity lead/liquid-argon
(LAr) sampling electromagnetic calorimeters covering the pseudorapidity range $\abseta<3.2$.
A steel/scintillator tile calorimeter provides hadronic energy measurements in the pseudorapidity region
$\abseta<1.7$. In the regions $1.5<\abseta<4.9$ the hadronic energy measurements are provided by two 
endcap LAr calorimeters using copper or tungsten as absorbers. The calorimeters are surrounded 
by a muon spectrometer that provides muon tracking beyond the calorimeters in the range $|\eta|<2.7$, and improves 
muon momentum resolution, charge measurements, and identification including triggering. 
\par Events are selected using a three-level trigger system~\cite{PERF-2011-02}. A hardware-based Level-1 trigger uses a subset
of detector information to reduce the event rate to 75 kHz or less. The rate of accepted events
is then reduced to about 400 Hz by two software-based trigger levels, Level-2 and the Event Filter.
These events are then stored for later offline reconstruction and analysis.

\section{Data and simulated event samples}
\label{sec:dataMCsamples}
\par This analysis uses a data set of $pp$ collisions collected 
at a center-of-mass energy $\sqrt{s} = 8$~\TeV during 2012 under stable 
beam conditions. After applying data quality requirements, the data set has 
a total integrated luminosity of $20.2 \pm 0.4~\ifb$~\cite{newLumiPaper}.
\par The exclusive SM $\yyWW$ signal sample
is generated using the \HERWIGPP~\cite{Hpp} Monte Carlo (MC) generator, while  $\yyWW$ 
signal samples with both the SM and non-SM aQGC predictions are generated by \FPMC~\cite{fpmc}. 
These two generators use the EPA formalism with a standard dipole 
parameterization~\cite{Budnev:1974de} of the proton electromagnetic form factors 
to produce an equivalent photon flux in $pp$ collisions.
\FPMC is used in these studies to generate $pp\rightarrow
pggp\rightarrow pHp$ events. 
None of these exclusive $\WW$ and Higgs boson generators support the case where one or both of the initial protons dissociate.
\par Produced via a mechanism similar to that for the exclusive $\WW$ signal, exclusive $\tau^+\tau^-$ production is an irreducible 
background when the two $\tau$ leptons decay to an $\emu$ final state.
Elastic \yytautau, \yymumu\, and \yyee\ backgrounds are generated 
using \HERWIGPP. Single- and double-dissociative \yymumu\ and \yyee\ backgrounds are produced using 
\LPAIR~4.0~\cite{Vermaseren}, while \PYTHIA8~\cite{Pythia8} is used to produce single-dissociative \yytautau\  candidates.
Double-dissociative $\yytautau$ samples are not available but their contribution is small.
This paper refers to the $\tau$ processes described in this paragraph as the exclusive background. 
In the exclusive Higgs boson search, exclusive $\WW$ production is an additional background.
\par Inclusive $\WW$ production is a dominant background and has similar final states 
to the signal process, except that it is usually accompanied by additional
charged particles from the underlying event. 
The inclusive $\WW$ background is the sum of nonresonant $q\bar{q}\rightarrow \WW$ events, $gg \to \WW$ events 
from nonresonant direct production, and resonant production and decay of the 125~\GeV Higgs boson. 
The $q\bar{q}\rightarrow \WW$ and $H \rightarrow \WW$ samples are generated using the 
\POWHEG-\textsc{Box}~\cite{PowhegBox, Alioli:2008tz,Powheg0,Powheg1,Powheg2} generator (hereafter referred to as \POWHEG) interfaced to 
\PYTHIAeight (\PowhegPythiaEight) for parton showering, hadronization, and underlying-event simulation. 
The AU2~\cite{ATLAS:Pythia8Tune} parameter set (``tune'') is used for the underlying event.
For the nonresonant $gg \to \WW$ sample, the \ggww~\cite{gg2WW} program is used and the showering, 
hadronization, and underlying event
are simulated using \HERWIG~\cite{Herwig6.5} and \JIMMY~\cite{Jimmy}, with the AUET2~\cite{atl-phys-pub-2011-009} tune.
The CT10 PDF set \cite{CT10} is employed for all of these samples. 
The contribution from vector-boson fusion production of $\WW$ events,
generated  with \SHERPA~\cite{Sherpa} with CT10 PDFs, 
 is also included.
In all regions of phase space, a normalization factor of 1.2 
is applied to inclusive \WW\ background as a correction to the
cross-section as described in Section~\ref{sec:higgsControlRegions}.
\par Other backgrounds such as $W/Z$+jets are easier to reject than inclusive $\WW$ production,
because, in addition to being produced with extra charged particles, their final state topologies are
also different. However, their contribution is non-negligible due to their several orders of
magnitude higher cross-section. 
Both $W/Z$+jets processes are modeled with \ALPGEN~\cite{Alpgen} 
interfaced to \PYTHIAsix~\cite{Pythia6} (\AlpgenPythiaSix) using the
CTEQ6L1 PDF set~\cite{CTEQ6L1} and Perugia 2011C~\cite{PerugiaTune} tune. Diboson processes such as 
$WZ$ and $ZZ$~\footnote{The symbol $Z$ in $WZ$ and $ZZ$ is used here for both $Z$ and $\gamma^{*}$ production 
decaying to a lepton pair.} are also sources of background 
if exactly two charged lepton candidates are reconstructed and identified.
The $WZ$ and $ZZ$ samples are generated using {\PowhegPythiaEight}~\cite{Melia:2011tj} with 
the AU2 tune and the CT10 PDF set. 
Other diboson processes ($W\gamma$ and $Z\gamma$) are also considered, but their contributions are found to be negligible.
The \POWHEG generator interfaced to \PYTHIAsix with the CT10 PDF set is used to simulate $t\bar{t}$ background. 
Single-top-quark production through the $t$-channel is modeled with
\ACERMC~\cite{Kersevan:2004yg} interfaced to \PYTHIAsix with the CTEQ6L1 PDF set, 
while $s$-channel and $Wt$ single-top-quark backgrounds are simulated
using \MCatNLO~\cite{MCatNLO} interfaced to \HERWIG and \JIMMY with the CT10 PDF set and AUET2 tune.
The underlying event AUET2B~\cite{atl-phys-pub-2011-009} tune is employed for the $t\bar{t}$ and $t$-channel single-top-quark backgrounds.
A summary of the processes and simulation tools used in this paper are
given in Table~\ref{tab:MCSamples1}.
\par The same background samples are used for the exclusive Higgs boson search, except for $Z$+jets, 
which is modeled with \ALPGEN interfaced to \HERWIG and \JIMMY (\AlpgenHerwig) and top-quark background whose 
contribution to the exclusive Higgs boson signal region is negligible. 
The CTEQ6L1 PDF set is employed for the $\AlpgenHerwig$ $Z$+jets samples.
Two more sets of $Z$+jets samples, generated using {\PowhegPythiaEight}
and \SHERPA with CT10 PDF set, are used for additional background studies. 
\par All the background samples mentioned above are processed through a simulation of the ATLAS detector~\cite{ATLAS:Simulation} 
based on GEANT4~\cite{geant4}. The signal samples are processed through the fast detector simulation program ATLFAST2~\cite{PUB-2010-013}.
The effect of the multiple $pp$ collisions, which is 
referred to as pileup throughout this paper, is also simulated by overlaying minimum-bias events generated using \PYTHIA8 and corrected to agree with data.
\begin{table}[!h]
 \begin{center}
  \resizebox{\textwidth}{!}{
   \begin{tabular}{lr}
    \hline\hline
    Process 										         & MC Generator            \\
    \hline
    Exclusive $\WW$ signal\\
      \phantom{AAA} $\gamgamWWll$ ($\ell, \ell' = e,\mu,\tau$)                       & \HERWIGPP         \\
     aQGC signal            \\
       \phantom{AAA}$\gamgamWWll$ with $a_{0,C}^W/\Lambda^2 \neq 0$ 	      & \FPMC             \\
    Exclusive Higgs boson signal \\
        \phantom{AAA} Exclusive \ggHWWll				  & \FPMC             \\
     \\
	Exclusive dilepton  \\ 
        \phantom{AA} $\yyll$ ($\ell = e,\mu,\tau$) &  \HERWIGPP, LPAIR, \PYTHIAeight       \\
	{Inclusive $\WW$}     \\ 
        \phantom{AA} $W^+W^- \to \ell^{+} \nu \ell'^{-} \bar{\nu}$ ($\ell, \ell' = e,\mu,\tau$)     & \PowhegPythiaEight,\ $\ggww$+\HERWIG      \\
		\phantom{AA} Inclusive \ggHWWll                                	   & \PowhegPythiaEight              \\
        \phantom{AA} Vector-boson fusion $W^+W^- \to \ell^{+} \nu \ell'^{-} \bar{\nu}$        & \SHERPA       \\
	{Non-$\WW$ diboson (Other-$VV$ diboson)}  \\
		\phantom{AA} $WZ$, $ZZ$ 						& $\PowhegPythiaEight$  	    		\\
	{Other background}   \\ 
		\phantom{AA} $W$ + jets 									& \AlpgenPythiaSix																\\
		\phantom{AA} $Z$ + jets									& \AlpgenPythiaSix,\ \AlpgenHerwig			\\
		\phantom{AA} \ttbar, single top-quark, $Wt$				& $\POWHEG$+\PYTHIAsix,\ $\ACERMC$+\PYTHIAsix,\ $\text{\MCatNLO}$+\HERWIG          \\
		\hline        
   \end{tabular}
  }
 \end{center}
 \caption{A list of the simulated samples used for estimating the expected contributions 
          to the exclusive $\WW$ signal region and exclusive Higgs boson signal region.
		  The exclusive $\WW$ production is treated as background in the exclusive Higgs boson channel.
          Similarly, the exclusive Higgs boson production is a background to exclusive $\WW$ signal.}
 \label{tab:MCSamples1}
\end{table}
\section{Selection of leptons, jets, and charged particles}
\label{sec:physicsObjects}
\par Selection criteria are applied to the data and simulated samples to identify events that have good quality electron and muon candidates.
Electron candidates are reconstructed from clusters of energy deposited in the electromagnetic calorimeter
that are matched to tracks in the inner detector. They are required to
have transverse momentum $\pT > 10~\GeV$, and be within a pseudorapidity range
$\abseta<2.47$, excluding the region $1.37\leq\abseta\leq1.52$. Also,
they satisfy shower shape and track selection criteria that make up the ``very tight'' likelihood criteria~\cite{ATLAS-CONF-2014-032}
defined by a multivariate likelihood algorithm.
Electrons are required to be isolated based on tracking and
calorimeter information. Efficiencies for very tight electron 
identification range from $60\%$ to $70\%$.
Muon candidates with $\pT>10~\GeV$ are reconstructed from tracks in the inner detector matched to tracks in the muon
spectrometer. Muon candidates are required to be within a pseudorapidity range $\abseta<2.5$ and must 
satisfy the criteria outlined in Ref.~\cite{PERF-2014-05}, providing muon identification efficiencies 
of up to 95\%. 
The tracking and calorimeter isolation criteria for muon and electron candidates are the same as those used in Ref.~\cite{ATLAS:2014aga}.
\par 
Jets with $|\eta|<4.5$ are reconstructed from energy clusters in the calorimeter  using 
the anti-$k_t$ algorithm~\cite{Cacciari200657} with a radius parameter of 0.4. 
To suppress jets from pileup, only jets with $\pT>25$~GeV are considered. 
Missing transverse momentum $\boldsymbol{p_{\mathrm
T}^{\mathrm{miss}}}$ with magnitude 
$\MET$ is reconstructed as the magnitude
of the negative vector sum of 
the momentum of reconstructed physics objects -- $e$, $\mu$, photons, and jets -- 
and remaining calorimeter clusters that are not associated with any hard
objects are also included with the proper calibration~\cite{PERF-2011-07}. 
\par Charged particle tracks having $\pT>0.4$~\GeV and $|\eta|<2.5$ reconstructed 
by the inner detector are used in this paper to reject nonexclusive production. 
They are required to leave at least one hit in the pixel detector and 
at least four hits in the silicon microstrip detector.

\section{Exclusivity selection}
\label{sec:exclSelection}
Exclusive candidates are characterized by large rapidity gaps~\cite{Aad:2015xis, RapGaps}
 between the protons and the system of interest -- a $\WW$ pair or 
Higgs boson. A signature for this, in the ATLAS detector, is an absence of tracks, other than tracks from 
the \WW\ pair or Higgs boson decay products.  
Inclusive candidates, in contrast, are produced with extra particles that
originate from the emission and hadronization of additional gluons, and the underlying event. 
These extra particles usually produce tracks in the inner detector. This
analysis takes advantage of the absence of additional charged particle
tracks to separate exclusive from inclusive (color processes) production.
\par In exclusive Higgs boson and $\WW$ production, no further
 charged particles are produced apart from 
the two final-state leptons. 
So in order to select exclusive events, the distance between the $z_0$ of the leptons is 
required to be less than 1~mm, where $z_0$ is the $z$-coordinate at the point of closest 
approach of a lepton (or track) to the beam line in the $r$-$\phi$ plane. Then the average $z_0$ of
the two leptons, $z_0^{\mathrm{av}}$, is taken as the event vertex
and is referred to as the \emph{lepton vertex}.
In this paper, an \emph{exclusivity selection}
is applied, which requires zero additional tracks with $\pT > 0.4$ GeV
near $z_0^{\mathrm{av}}$ 
with $|z_0^{\mathrm{track}}-z_0^{\mathrm{av}}| < \DZ$. To improve the efficiency for exclusive events whose 
leptons have more than one associated track (due to bremsstrahlung for example), 
candidate tracks considered for this selection are required to be unmatched to either of the final-state leptons. 
Therefore, a candidate track within an angular distance $\Delta R < 0.01$ and within 1~mm in \zzero of
either of the final-state leptons is considered matched and is ignored. 
The value $\DZ$ is optimized
using exclusive Higgs boson and exclusive $\WW$ simulated samples. 
A value of $\DZ=1$ mm is chosen for all results in this paper.
The exclusivity selection efficiency is found to be 58\% and is largely process
independent as is discussed in Section~\ref{sec:photonFlux}.
In Fig.~\ref{fig:pileupEff} the exclusivity efficiency is extracted from  exclusive Higgs boson signal 
simulated by \FPMC, plotted against the average number of interactions per beam crossing $\mu$.
For the dataset used in this study $\left< \mu \right>$ is 20.7.
\begin{figure}
\centering
 \includegraphics[width=0.7\linewidth]{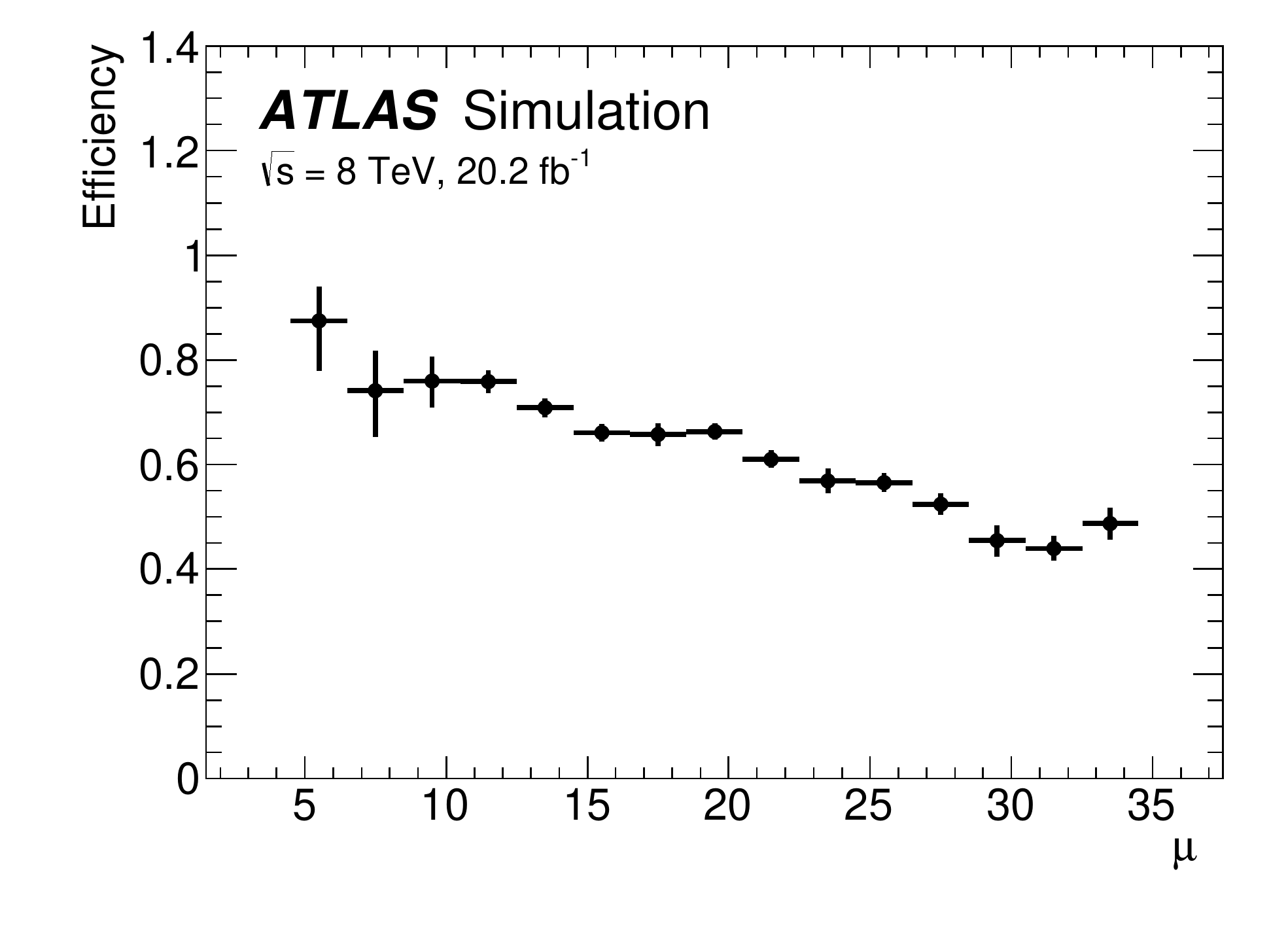} 
\caption{Efficiency of the exclusivity selection, extracted from the
exclusive Higgs boson signal simulation, is plotted 
against the average number of interactions per beam crossing
$\mu$. The average is 20.7 for the current dataset.}
\label{fig:pileupEff}
\end{figure}

\section{Event selection}         
\label{sec:eventSelection}
\par Events are required to satisfy at least one of  the
single- and dilepton triggers in 
Table~\ref{tab:triggers}. They are further categorized into $ee$, $\mu\mu$, and $e\mu$ final states. 
A combination of single-lepton and different-flavor dilepton triggers is used to select 
the signal events, while the same-flavor dilepton triggers are used to select $ee$ and 
$\mu\mu$ events for validation and control regions.
\par For both the exclusive $\WW$ and Higgs boson channels, this analysis selects candidates
consistent with leptonic decays of $W$-boson pairs into oppositely
charged different-flavor
leptons. Additional kinematic requirements reject background while
retaining as much of the signal as possible. 
Exclusive $\WW$ production is a large background in the exclusive Higgs boson search, 
while the exclusive Higgs boson contribution to the exclusive $\WW$ signal is negligible.
So the kinematic requirements for the two channels differ slightly.
Table~\ref{tab:selections} summarizes the selection criteria for both channels.
\begin{table}[!h]
 \centering
 \begin{tabular}{lr}
  \hline\hline
 Trigger  								& Lepton \pT\ criteria [\GeV] \\
\hline
 Single electron   	& $\pT^{e} > 24$ \\
 Single muon  			& $\pT^{\mu} > 24$ \\
 Symmetric dielectron   	& $\pT^{e_1} > 12$,\ \  $\pT^{e_2} > 12$\\
 Asymmetric dimuon      	& $\pT^{\mu_1} > 18$,\ \  $\pT^{\mu_2} > 8$ \\
 Electron-muon  	& $\pT^{e} > 12$,\ \ $\pT^{\mu} > 8$ \\
  \hline
\hline
 \end{tabular}
\caption{Single-lepton and dilepton triggers are used to select event candidates.
         Single-lepton triggers require either of the leptons to satisfy the 
         specified $\pT$ criterion, while dilepton triggers have two specific 
         $\pT$ criteria.}   
\label{tab:triggers}
\end{table}
\subsection{Exclusive $\WW$ candidate selection}
\label{sec:eventSelectionWW}
\par For the exclusive $\WW$ channel, requiring oppositely charged $e^{\pm}\mu^{\mp}$ leptons
rejects same-flavor lepton events from Drell-Yan and exclusive dilepton processes.
The invariant mass of the dilepton system is required to be greater 
than $20~\GeV$. This rejects a significant fraction of the remaining 
background in which jets have non-prompt or fake electron and/or muon signatures. 
The lepton with the higher $\pT$ is referred to as the leading lepton ($\ell 1$) 
and the other, the subleading lepton ($\ell 2$). 
The $\pT$ requirement on the leading lepton is chosen to be higher 
than the single-lepton trigger threshold, resulting in different leading and subleading 
leptons requirements: $\pT^{\ell 1} > 25~\GeV$ and $\pT^{\ell 2} > 20~\GeV$, respectively. 
These selection criteria define
$preselection$.
\par To reduce $\yytautau$ and $\DrellYantotautau$
contamination, the magnitude of the transverse momentum of the
dilepton system  ($\pT^{e\mu}$) 
is required to be greater than 30~\GeV. The exclusivity 
requirement rejects most of the remaining inclusive background. 
After applying these selection criteria, 70\% of the predicted background is 
due to inclusive $\WW$ production, while $\yytautau$ contributes 15\% and the
contributions from other categories are negligible.
\par The limits on aQGCs are extracted from the region with $\pT^{e\mu} > 120~\GeV$. This requirement 
considerably reduces the SM contribution.
\subsection{Exclusive Higgs Boson candidate selection}
\label{sec:eventSelectionHiggs}
\par The Higgs boson decays to $\WW$ give one on-shell and one off-shell $W$ boson. Thus the subleading lepton 
minimum $\pT$ is lowered to 15~\GeV. For the same reason, the $\memu$ threshold is lowered to 10~\GeV.
The other requirements in the 
preselection are the same as for the exclusive $\WW$ sample. In contrast to the $\WW$ topology, the zero spin 
of the Higgs boson implies that the final-state leptons have small
angular separation. 
Therefore the angular separation of the leptons in the transverse
plane (\dFem\ ) and the dilepton mass ($\memu$) are two  
good discriminating variables against the remaining exclusive $\WW$ background, which has a 
wider angular separation and relatively higher dilepton mass. 
Thus, \memu\ and \dFem\ selection criteria are further imposed in the Higgs boson search. 
The transverse mass of the Higgs boson system, $\mT$, is defined as:
\begin{equation}
  \label{eq:mT}
  \mT = \sqrt{(E_{\mathrm T}^{e\mu}+\met)^{2} - |\boldsymbol{\pTemu} + \boldsymbol{ p_{\mathrm T}^{\mathrm{miss}}}|^{2}},
\end{equation}
where $E_{\mathrm T}^{e\mu} = \sqrt{|\boldsymbol{\pTemu}|^{2}+m_{e\mu}^{2}}$ and $|\boldsymbol{ p_{\mathrm T}^{\mathrm{miss}}}|=\met$. 
Requiring $\mT<140$~\GeV further reduces both the inclusive and exclusive $\WW$ backgrounds and  
improves the signal significance by 20\% (see Fig.~\ref{fig:mTSig}).
The exclusivity selection uses \DZ = 1~mm here
as well.
\begin{table}
\renewcommand*{\arraystretch}{1.4} 
\centering
\begin{tabular}{l|c|c}
\hline\hline
                                &  $\WW$ selection                       & Higgs boson selection \\
\hline\hline
\multirow{3}{*}{\textit{Preselection} } & \multicolumn{2}{c}{Oppositely charged $e\mu$ final states}   \\
                                \cline{2-3}
                                & $\pT^{\ell 1} > 25~\GeV$ and $\pT^{\ell 2} > 20~\GeV$  & $\pT^{\ell 1} > 25~\GeV$ and $\pT^{\ell 2} > 15~\GeV$ \\
                                \cline{2-3}
								& $\memu > 20~\GeV$                                     & $\memu > 10~\GeV$     \\
\hline
\multirow{3}{*}{\ }				& \multicolumn{2}{c}{$\pTemu > 30~\GeV$}\\
                                \cline{2-3}
                                & \multicolumn{2}{c}{Exclusivity
                                selection, \DZ}\\
\hline
aQGC signal                     & $\pTemu > 120~\GeV$ & -- \\
\hline
\multirow{3}{*}{Spin-0 Higgs boson}   &          --                                           & $\memu <55~\GeV$      \\
							    &          --                                           & $\dFem < 1.8$       \\
                                &          --                                           & $\mT <140$~\GeV       \\
\hline\hline
\end{tabular}
\caption{Selection criteria for the two analysis channels.}  
\label{tab:selections}
\end{table}

\section{Pileup and exclusivity validation with $\yyll$ events}
\label{sec:validationDilepton}
\par The selection strategy described in Section~\ref{sec:exclSelection}
 represents a new approach to extract exclusive processes without using
 usual vertex reconstruction~\cite{ATLAS-CONF-2012-042}. 
 This section describes two studies designed to validate this technique. 
 The first one demonstrates how the $\DZ$ selection gives results
 comparable to those of 
 previous strategies employed by the ATLAS Collaboration
 in a related measurement at $\sqrt{s} = 7~\TeV$~\cite{ATLAS7}, and the second one shows how 
 simulation of pileup and modeling of underlying event activity are verified.
 Except for possible non-standard couplings, the exclusive production of $\WW$ and $\ell^+\ell^-$ are similar. 
 Exclusive dilepton candidates are therefore used in both studies  
 because elastic $\yyll$ production can be separated from SD and DD production using dilepton transverse momentum, 
$\pT^{\ell\ell}$, and acoplanarity ($1-\nolinebreak |\Delta\phi_{\ell\ell}|/\pi$) of the dilepton system, where $\Delta\phi_{\ell\ell}$ 
is the dilepton azimuthal separation. The $\yymumu$ candidates are used for these studies 
while $\yyee$ candidates are used for cross-checks.
\par First, a measurement is made of the correction factor, \fEL, defined as
the ratio of observed elastic $\yymumu$ candidates to the $\HERWIGPP$ prediction
based on the EPA formalism. This factor is expected to be
lower than 1.0 due to the finite size effects of the
proton~\cite{DyndalProtonSize}. Alternative formulations give similar results~\cite{HKR}.
Candidates are required to have two muons with $\pT^{\mu} >
20~\GeV$, invariant mass $45 < \mmumu < 75~\GeV$ or
$\mmumu>105~\GeV$ and pass the exclusivity selection ($\DZ=1~\mm$). 
The Drell-Yan $\DrellYantomumu$ process is the dominant background while contributions from other backgrounds are negligible.
The elastic $\yymumu$ signal is enhanced by selecting the low-$\pTll$ region 
with an upper limit on $\pTll$ varied between 3~\GeV to 5~\GeV to
study systematic uncertainties. 
\par The value of \fEL\  
is extracted from template fits in acoplanarity. 
Some of the contributing processes have similar acoplanarity shapes; in particular
the Drell-Yan and DD backgrounds are not distinguishable.
Two fitting strategies are pursued. The first template strategy
attempts to distinguish 
three shapes: elastic, SD, and combined DD plus background. The relative 
weighting of DD and background is varied to estimate the associated
systematic uncertainty. The second template
strategy uses the 
elastic and combined SD and DD shapes, with the background yield constrained to the simulation's 
prediction. These two fitting strategies give consistent results and 
 are stable at the level of $10\%$ under the variation of $\pTmumu$ and $\DZ$ selections,
 the four different Drell-Yan
 generators, bin width, 
and fit range. These variations reflect
mismodeling of $\pTmumu$ and systematic uncertainties related to shape correlations
and signal strength. The effect of these variations is much larger
than the $3\%$ combined effect of the systematic uncertainties discussed
in Ref.~\cite{ATLAS7}, which can then be ignored. 
The best-fit value is 
$\fEL = 0.76 \pm 0.04 (\textrm{stat.}) \pm 0.07 (\textrm{sys.})$,
where the systematic uncertainty covers the spread of fit
values, and Fig.~\ref{fig:dilepAcoFit3Shape} shows the acoplanarity
distribution compared to SM expectation normalized by the factors
determined in this fit.
An additional uncertainty of $10\%$ related to pileup is discussed in the following paragraph.
A similar study using $\yyee$ candidates yields a consistent correction factor but 
with lower precision; thus the final value for \fEL\ is taken from the
$\yymumu$ sample.  This correction factor is used
to correct
the number of $\yytautau$ candidates predicted 
by simulation in both the exclusive $\WW$ and the exclusive Higgs
boson signal regions. Similar suppression is
expected~\cite{DyndalProtonSize} and observed~\cite{ATLAS7} in
dissociative events, so the \fEL\ factor is applied to dissociative events as well.
\begin{figure}
\centering
 \includegraphics[width=0.5\linewidth]{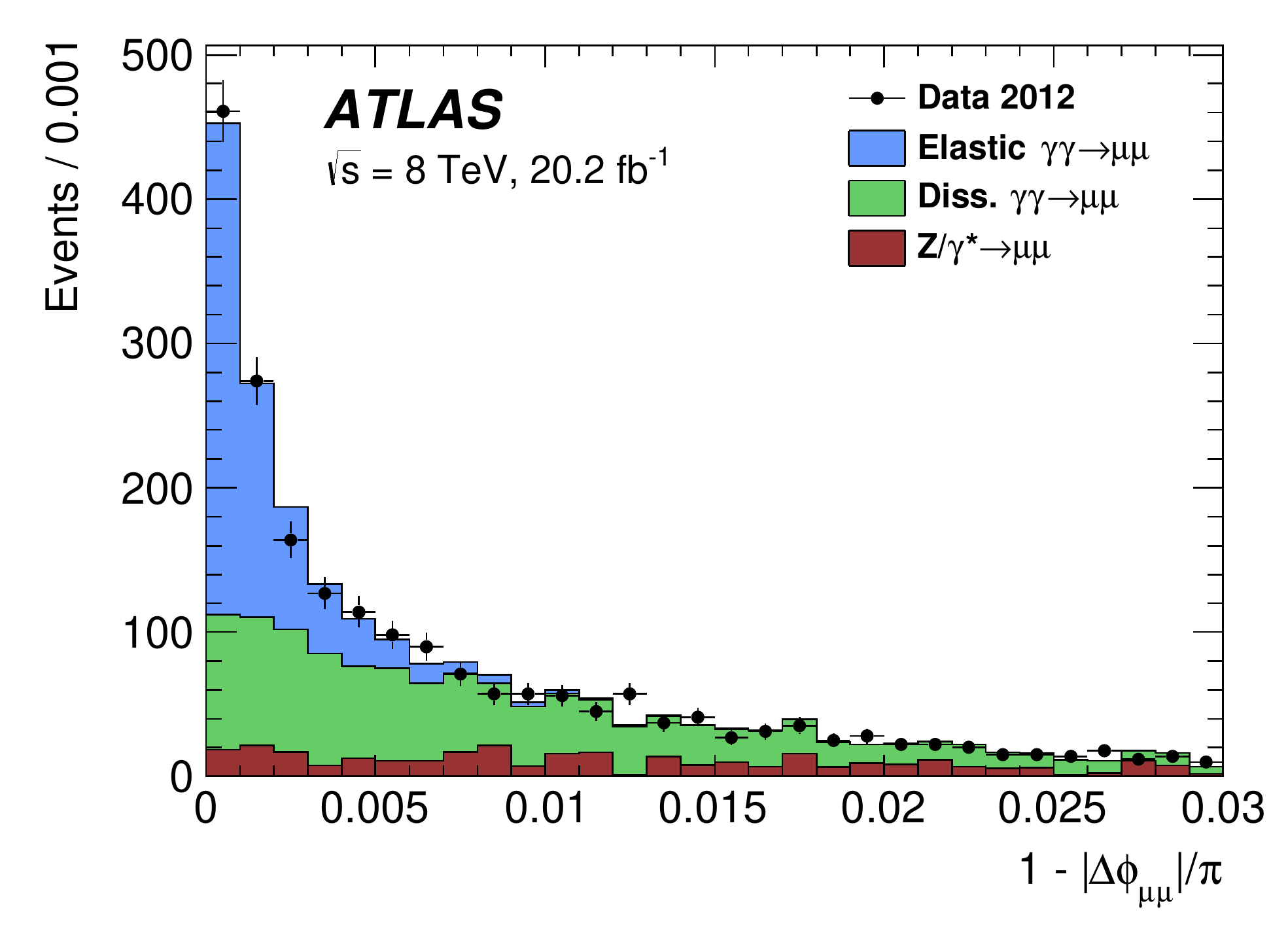} 
 \caption{Dimuon acoplanarity distributions after applying the
 exclusivity selection and requiring $\pT^{\mu\mu} < 3$~\GeV. The 
 expected Drell-Yan shape and the 
 elastic and combined SD and DD (Dissociative) shapes normalized from
 the fit are stacked. This fit 
          determines the factor $\fEL$.}
 \label{fig:dilepAcoFit3Shape}
\end{figure}
\par In the second study, the impact of pileup on the signal efficiency
 and accuracy of the modeling in the simulation
 is evaluated. A kinematic selection is defined to enhance
 the fraction of elastic events. Events with $\pTmumu < 3~\GeV$ and
 acoplanarity $<$ 0.0015 are studied with both the nominal exclusivity selection criteria and by
 demanding exactly one extra track within $\DZ = 3~\mm$. In the
 case of exclusive signal, when there is one extra track, the extra
 track is from
 pileup and its $\Delta z_0=|z_0^{\mathrm{track}}-z_0^{\mathrm{av}}|$ has a
 locally constant distribution 
 while  for any inclusive
 background  the track
 originates from the same vertex and the $\Delta z_0$ distribution
 peaks at zero as can be seen in
 Fig.~\ref{fig:dzEventOneTrk}.  A normalization factor, the
 background-subtracted ratio of observed exclusive events to the predicted
 sum of elastic, SD, and DD is determined for both selections.  For
 nominal (zero track) exclusivity this normalization factor 
 is $0.73 \pm 0.03 {\mathrm{(stat.)}} \pm 0.01 {\mathrm{(sys.)}}$.  The one-track selection,
 illustrated in Fig.~\ref{fig:dzEventOneTrk}, gives a factor of
 $0.70 \pm 0.06 {\mathrm{(stat.)}} \pm 0.03 {\mathrm{(sys.)}}$ where the systematic
 uncertainties result from the uncertainty in the background
 normalization factor. 
 The zero-track and one-track normalization factors are consistent at the
 level of $10\%$, which is taken to be a measure of the accuracy of
 the pileup simulation in predicting signal efficiency.
\par The value of \fEL with the additional $\pm 10\%$ relative systematic
 uncertainty for signal efficiency added in quadrature with the
 previous systematic uncertainty               
\begin{equation}                                                               
 \fEL = 0.76 \pm 0.04 (\textrm{stat.}) \pm 0.10 (\textrm{sys.})                
 \label{eqn:fELfactor}                                                         
\end{equation}                                                                 
is consistent with the value  
of $0.791 \pm 0.041~(\mathrm{stat.}) \pm 0.026~(\mathrm{sys.}) \pm
0.013~(\mathrm{theory})$ obtained in an earlier analysis using data
from $pp$ collisions at $\sqrt{s}$ = 7~\TeV~\cite{ATLAS7}.
This value is also consistent with the theoretical estimate 
of \fEL $\sim$ 0.73--0.75, related to the proton size effects in the probed region of
dimuon mass~\cite{DyndalProtonSize}. 
\begin{figure}
\centering
 \includegraphics[width=0.5\linewidth]{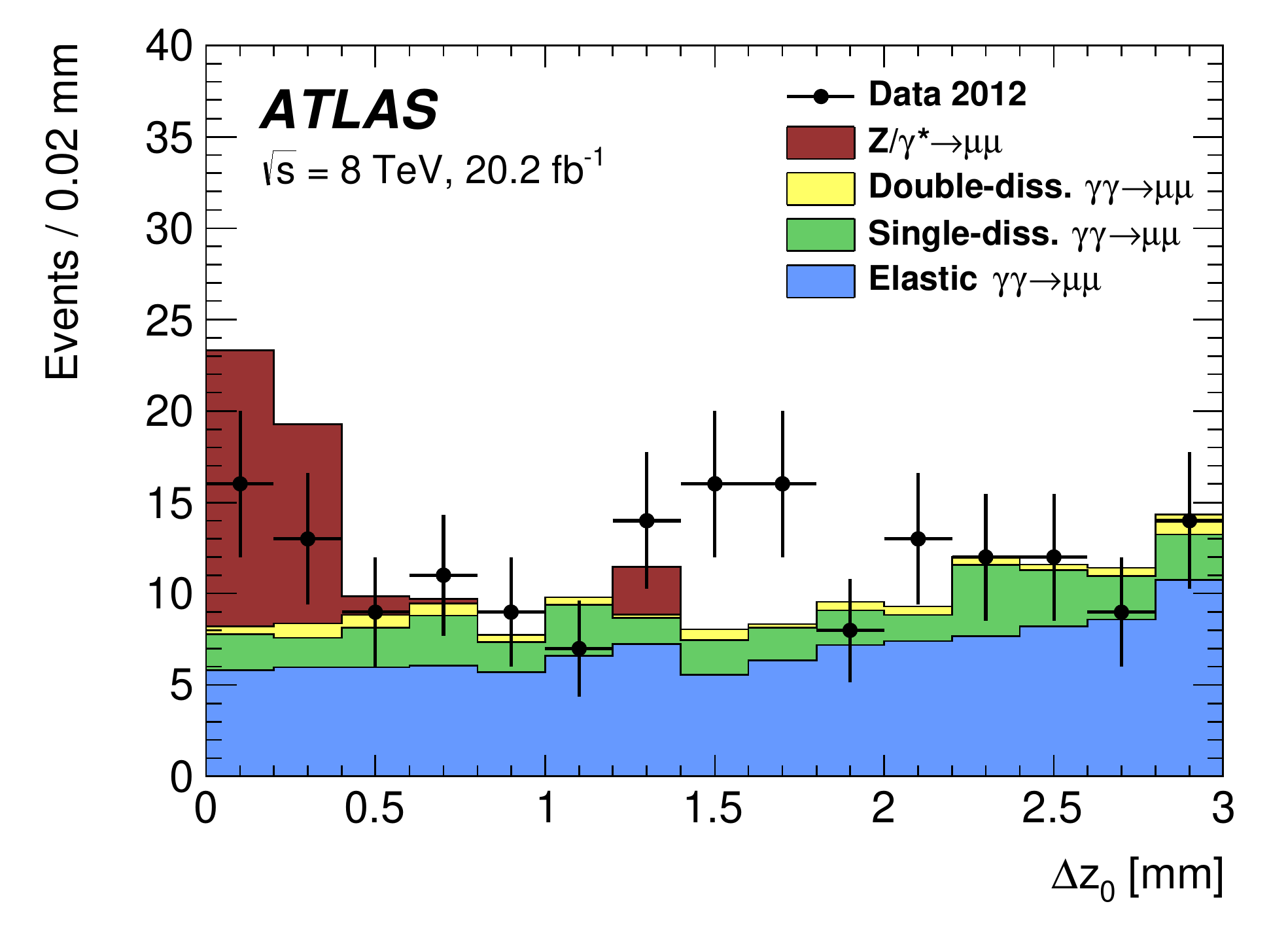} 
 \caption{Absolute $\Delta z_0$ of the extra track to the lepton vertex 
          in the region defined by acoplanarity $<0.0015$. The exclusivity 
          requirement was changed to select exactly one
          extra track within 3~mm.  The exclusive predictions are
 scaled by a factor of 0.70.}
 \label{fig:dzEventOneTrk}
\end{figure}

\section{Signal and background control regions}
\label{sec:controlRegions}
Several control regions are established to use data events to 
cross-check simulations in areas where they are known to be less 
reliable. The ratio of elastic to dissociative contributions is 
extracted from one control region, since a simulation for $\yyWW$ 
dissociative events is not available. Another set of control regions 
is used to study the proximity of small numbers of extra tracks to 
the lepton vertex. This is another regime where the underlying-event 
models have not been thoroughly tested, so relying on the data is preferred. 
Finally a control region is established for inclusive $\WW$ production, 
a predominant background. This control region has a different exclusivity 
requirement, one to four extra tracks, in order to increase the fraction of 
inclusive $\WW$ events.
The inclusive $\WW$ contribution to the exclusive $\WW$ signal region is 
estimated using a data-driven method. Based on the number of events observed
in this control region, this method makes some assumptions about the rejection 
of background when going from the control (one to four tracks) to the
nominal  (zero tracks)  exclusivity requirement, 
and derives an estimate for the background from inclusive $\WW$, Drell-Yan, 
$W$+jets, and top-quark production. The latter three processes have collectively a smaller 
contribution and are referred to as \textit{other background}. 
Other contributions to the background are derived from Monte Carlo simulation and 
are found to be negligible.
\subsection{Single-dissociative and double-dissociative contributions}
  \label{sec:photonFlux}
\par Without detecting the outgoing protons, the elastic $\yyWW$ events are indistinguishable 
from SD and DD candidates. However, simulations are only available for the elastic $\yyWW$ process;
predictions for dissociative production of $\WW$ are not available. 
Following the strategy in Ref.~\cite{CMSWW} a normalization
factor, $\fgam$, is determined.  This factor is used
to correct the prediction for elastic $\yyWW$ to account for dissociative
events. 
It is computed from data using $\yymumu$ candidates that satisfy 
the exclusivity selection with $\DZ = 1~\mm$, $\pT^{\mu} > 20~\GeV$ and 
$m_{\mu\mu} > 160~\GeV$ ($\sim2m_W$). 
The factor $\fgam$ is defined as the ratio of observed 
dimuons in data to the \HERWIGPP\ prediction for elastic dimuon production: 
\begin{equation}
 \fgam = \frac{N_{\mathrm{Data}} 
              - N^{\POWHEG}_{\mathrm{Background}}}{N^{\HERWIGPP}_{\mathrm{Elastic}}} \Bigg\rvert_{m_{\mu\mu} > 160~\GeV}
            = 3.30 \pm 0.22 \mathrm{(stat.)} \pm 0.06 \mathrm{(sys.)},
 \label{eqn:fgamma}
\end{equation}
where $N_{\mathrm{Data}}$ is the number of candidates in the data, 
$N_{\mathrm{Background}}^{\POWHEG}$ is the expected number of
background events, 
and $N_{\mathrm{Elastic}}^{\HERWIGPP}$ is the expected number of elastic $\yymumu$ candidates
directly from \HERWIGPP, i.e, the unscaled EPA prediction. 
Drell-Yan processes are the main sources of background, whereas
inclusive and exclusive $\WW$ processes
contribute less than $10\%$. The uncertainty is predominantly statistical, but
also contains a systematic component estimated by varying the
$\PowhegPythiaEight$ Drell-Yan correction factor 
by $\pm 20\%$, as is be discussed in Section~\ref{sec:MCTrackModelling}. 
Predictions for this ratio are becoming available~\cite{HKR1}.
\par The dilepton invariant mass distributions for the $\mu^+\mu^-$ and $e^+e^-$ final states are 
shown in Fig.~\ref{fig:fluxM160}. The elastic contribution is scaled by $\fEL = 0.76$ 
and the SD contribution is normalized so that the sum of the elastic and SD contributions corresponds 
to $\fgam \times N_{\mathrm{Elastic}}^{\HERWIGPP}$. 
The shapes of the SD and DD samples are quite similar, so
the SD shape is used to describe both the SD and DD 
processes. 
The data are well described by the simulation over the full mass
range. While the range of $m_{\ell\ell}>160~\GeV$ was chosen to
correspond to the threshold 
  $m_{WW} > 2m_W$, the value of $f_\gamma$ is in fact
  rather insensitive to the choice of this threshold. The $\WW$ sample tends
to have higher $m_{WW}$ than these dilepton control samples $m_{\ell\ell}$. The $m_{ee}$ distribution in 
Fig.~\ref{fig:fluxM160} shows that $\fgam$ is also valid for the
electron channel. Therefore, the total expected $\yyWW$ event yield 
in both the exclusive $\WW$ and the exclusive Higgs boson channels is taken
to be the product of
 $\fgam$ times the \HERWIGPP prediction for elastic $\yyWW$ production.
\begin{figure}
 \includegraphics[width=0.5\linewidth]{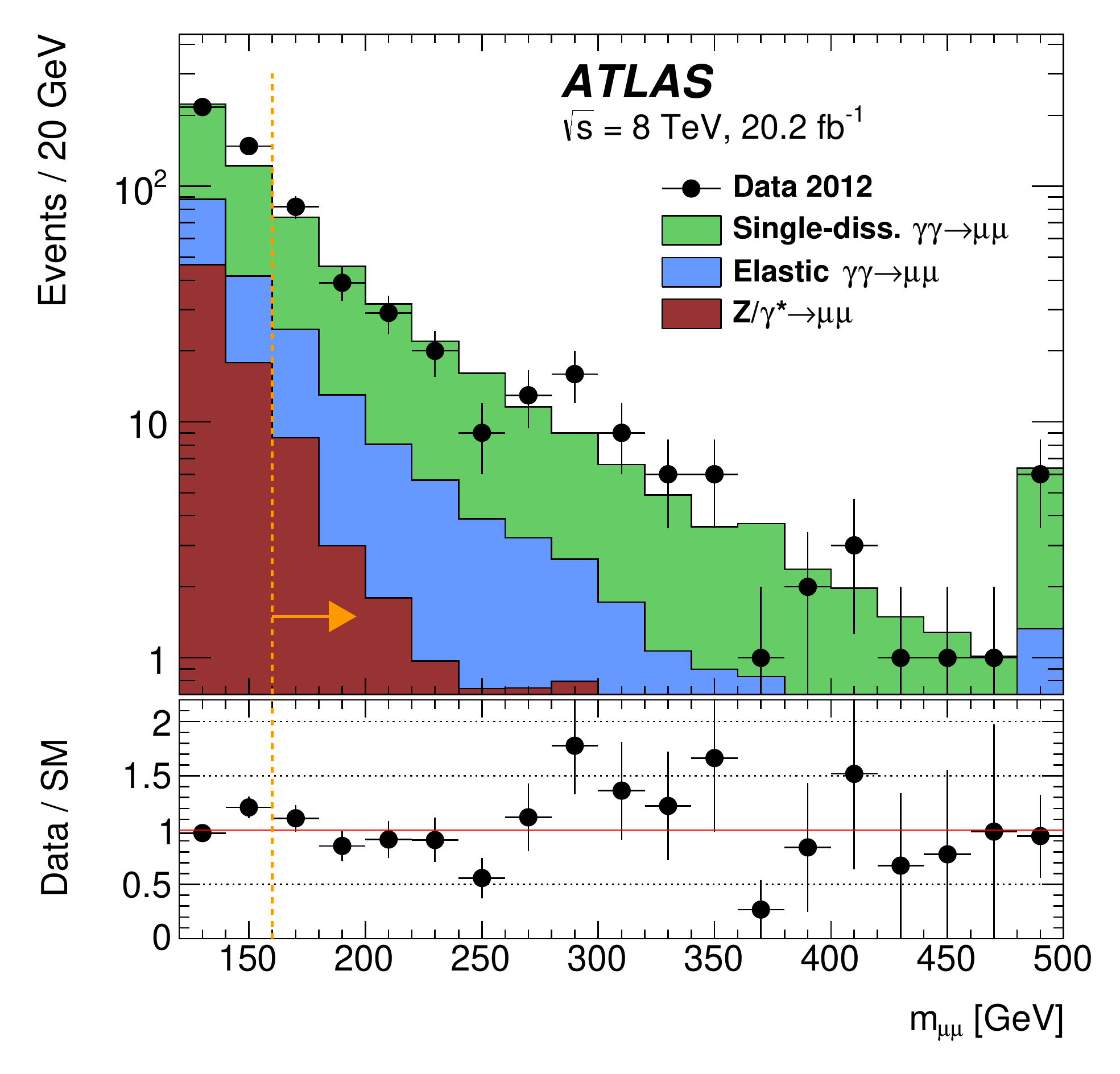} 
 \includegraphics[width=0.5\linewidth]{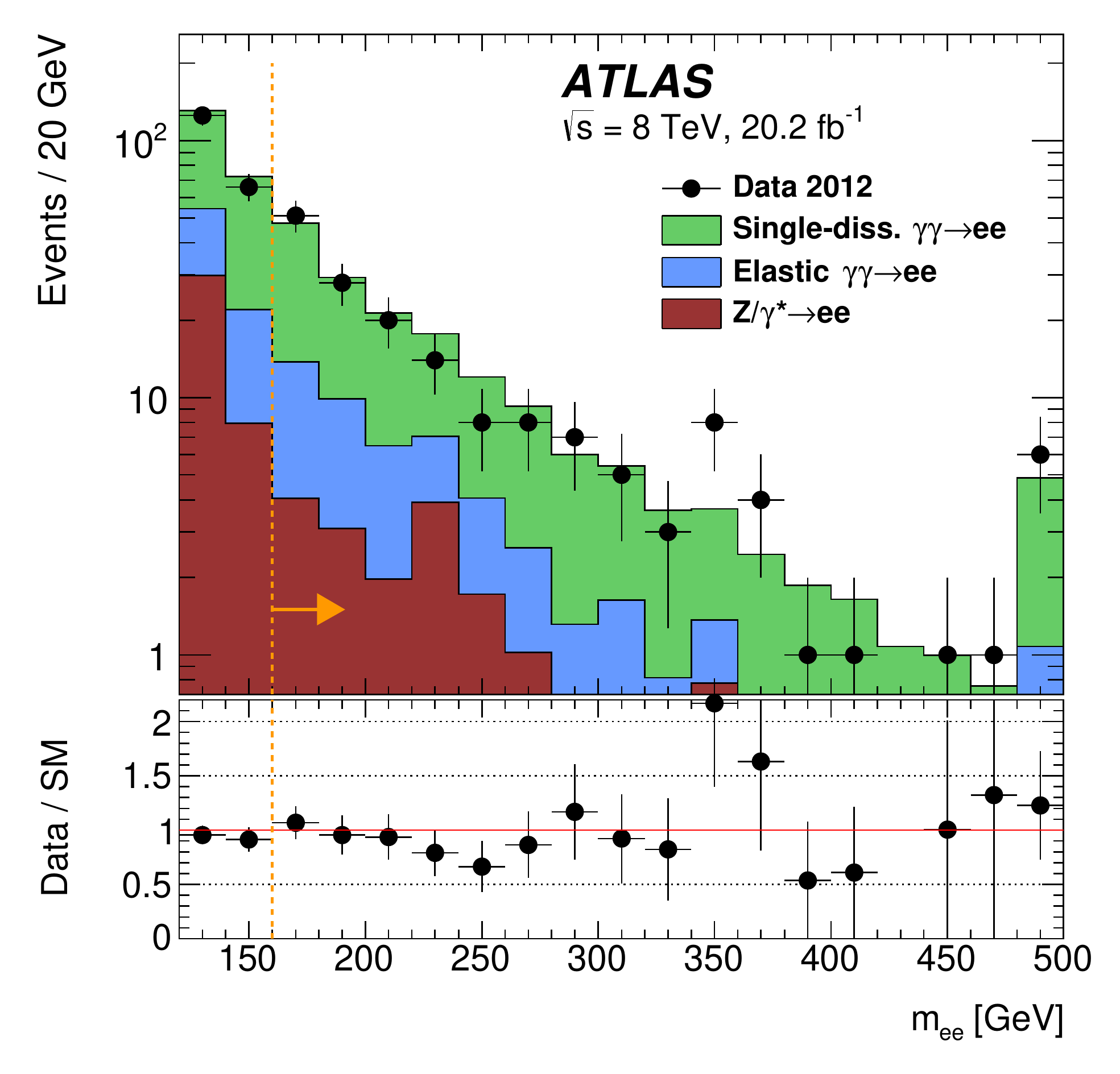} 
 \caption{The dilepton invariant mass distribution for muon
 candidates (left) and 
          electron candidates (right). The elastic yield is scaled 
          by $\fEL = 0.76$ and the SD distribution is scaled to
          bring the sum of the elastic and SD contributions to the 
          \HERWIGPP prediction for the elastic process multiplied by
          the $\fgam$ factor in the mass region above
          160~\GeV. The last bin includes overflow.}
 \label{fig:fluxM160}
\end{figure}
\par The dimuon signal sample with
 mass above 160~\GeV is also used to determine the signal efficiency
 for  exclusivity,
 which is $0.58 \pm 0.06$ where the 10\% uncertainty arises from pileup 
 modeling as described in 
 Section~\ref{sec:validationDilepton}. Other signal samples give compatible results.
\subsection{Track multiplicity modeling}
  \label{sec:MCTrackModelling}
\par In $pp$ collisions, inclusive Drell-Yan, $\WW$, 
$t\bar{t}$, and many other events are initiated by quarks or gluons.
Through hard
radiation and the accompanying underlying event, such events are
produced with several additional charged particles. The exclusivity
selection is designed to reject such inclusive candidates that have
additional tracks near the dilepton vertex.
To estimate inclusive backgrounds
from Drell-Yan production of $\tau^+\tau^-$ and inclusive $\WW$ production, the track
multiplicity modeling of low-multiplicity candidates is studied with
a
high-purity $Z$ boson sample and scaled with appropriate correction factors.
\par Drell-Yan candidates are selected by requiring exactly two muons
with
$\pT^{\mu} > 20~\GeV$ and |$\eta$| < 2.4, and satisfying $m_{\mu\mu} > 45~\GeV$. 
The $Z$-resonance region, $80 < m_{\mu\mu} < 100~\GeV$, is used to measure the
efficiency of the exclusivity selection in both the data and
simulation. The contributions from 
non-$Z$ processes are subtracted before and after the exclusivity
selection for both the data and simulated samples. This non-$Z$ contribution is estimated from the sideband regions $70 < m_{\mu\mu} < 80~\GeV$
and $100 < m_{\mu\mu} < 110~\GeV$. 
The efficiency of the exclusivity selection for inclusive $Z$ events in data
is found to be 0.004. This was compared to efficiencies for simulated
Drell-Yan samples from four generators: 
\AlpgenPythiaSix, \AlpgenHerwig, \PowhegPythiaEight, and \SHERPA.
In general, the exclusivity criterion rejects more $Z/\gamma\to\mu^+\mu^-$
candidates in the data than in the simulation. The study was repeated
for events with one to four additional tracks.
\par Correction factors are defined as the ratio of the exclusivity selection
efficiency in data to the one in the simulation. They are
reported in Table 4 and denoted
by $f^{\mathrm{sim}}_{\mathrm{nTracks}}$, where sim is
$\mathrm{P}$ for \PowhegPythiaEight, 
$\mathrm{AH}$ for \AlpgenHerwig, and $\mathrm{AP}$ for \AlpgenPythiaSix, and
$\mathrm{nTracks}$ is the
number of additional tracks.
These correction factors are used to scale the Monte Carlo prediction for the inclusive 
processes considered in the paper.  The background event tuning for
simulation of low multiplicity in $8~\TeV$ data is seen to vary widely.
\par The uncertainties in these correction factors are estimated from the
variation of the exclusive efficiency as a function of $m_{\mu\mu}$ of
the various generators. To check the consistency of the predictions of
evolution of
underlying event multiplicity as a function of mass, ratios of the predictions of the three
generators to the one by \SHERPA are listed in Table~\ref{tab:gencm}.
These are normalized 
such that the average over the full mass range is 1. The variations are
typically within $20\%$, which is taken as the systematic uncertainty in
extrapolating the $f^{\mathrm{sim}}_{\mathrm{nTracks}}$ correction factors.
\begin{table}[h]
 \begin{center}
  \renewcommand*{\arraystretch}{1.2} 
 \begin{tabular}{l|c|c|c}
 \hline
 Number of extra tracks & \PowhegPythiaEight $\fP_n$ & \AlpgenHerwig $\fAH_n$ & \AlpgenPythiaSix $\fAP_n$\\
 \hline
 $n = 0$  & 0.58 &  0.21 & 0.69 \\
 $n = 1$--$4$ & 0.88  &  0.39 &  0.85 \\
 \hline
 \end{tabular}
 \caption{Ratio of exclusivity efficiencies for $Z \to \mu\mu$ production in data and
simulation for different generators after sideband subtraction of nonresonant
contributions. The efficiency ratios
$f^{\mathrm{sim}}_{\mathrm{nTracks}}$ are shown for exclusive
 selection ($n = 0$) as well as for a
relaxed selection with one to four additional tracks ($n = 1$--$4$).}
 \label{tab:rejscale}
 \end{center}
\end{table}
\begin{table}
\begin{center}                                                                 
\begin{tabular}{l|c|c|c}                                                       
\hline                                                    
Mass [\GeV]  & \AlpgenHerwig & \AlpgenPythiaSix  & \PowhegPythiaEight  \\    
\hline         
 44--60  & $0.81 \pm 0.02$  & $0.84 \pm 0.03$ & $0.99 \pm 0.09$\\     
 60--90 & $1.04 \pm 0.02$  & $0.98 \pm 0.03$  & $1.01 \pm 0.02$\\     
 90--116 & $1.00 \pm 0.01$ & $1.02 \pm 0.02$  & $1.00 \pm 0.02$ \\  
 116--200 & $0.89 \pm 0.10$& $1.04 \pm 0.19$  & $0.76 \pm 0.10$\\    
\hline      
\end{tabular}                     
\caption{Ratio of the exclusivity selection efficiency in Drell-Yan
 $\mu^+\mu^-$
production as a function of dimuon mass of different generators to
\SHERPA. A common normalization factor is applied to each column to
obtain an average ratio of 1. Only statistical uncertainties 
are shown. The statistical uncertainty from \SHERPA is included and contributes $2.9\%, 0.8\%,
0.7\%$ and $5.7\%$ in the four mass regions. }
\label{tab:gencm}            
\end{center}             
\end{table}
\par 
To validate the correction factors $f^{\mathrm{sim}}_{\mathrm{nTracks}}$, 
an $\emu$ sample 
was defined. Figure~\ref{fig:extraTracks} (left) shows the distribution of
the number of additional tracks after applying the $\WW$ preselection as
defined in Table~\ref{tab:selections}.
Applying a relaxed exclusivity selection to select $\emu$ candidates with one to four 
extra tracks yields a sample that has low enough statistical uncertainties and is dominated by 
Drell-Yan events for $\pT^{e\mu} < 30~\GeV$ as illustrated in Fig.~\ref{fig:extraTracks} (right). 
Selecting $m_{e\mu} < 90~\GeV$ further rejects non Drell-Yan contamination as shown in Fig.~\ref{fig:xtrkZjetsMll}. 
The correction factor for \AlpgenPythiaSix Drell-Yan, computed in the
region defined by $\pT^{e\mu} < 30~\GeV$ and 
$m_{e\mu} < 90~\GeV$, is found to be $0.90 \pm 0.11$, in good
agreement with  $\fAP_{1-4} = 0.85$ found above for $Z \to \mu^+\mu^-$. 
\begin{figure}
\begin{minipage}{0.45\linewidth}
 \includegraphics[width=\linewidth]{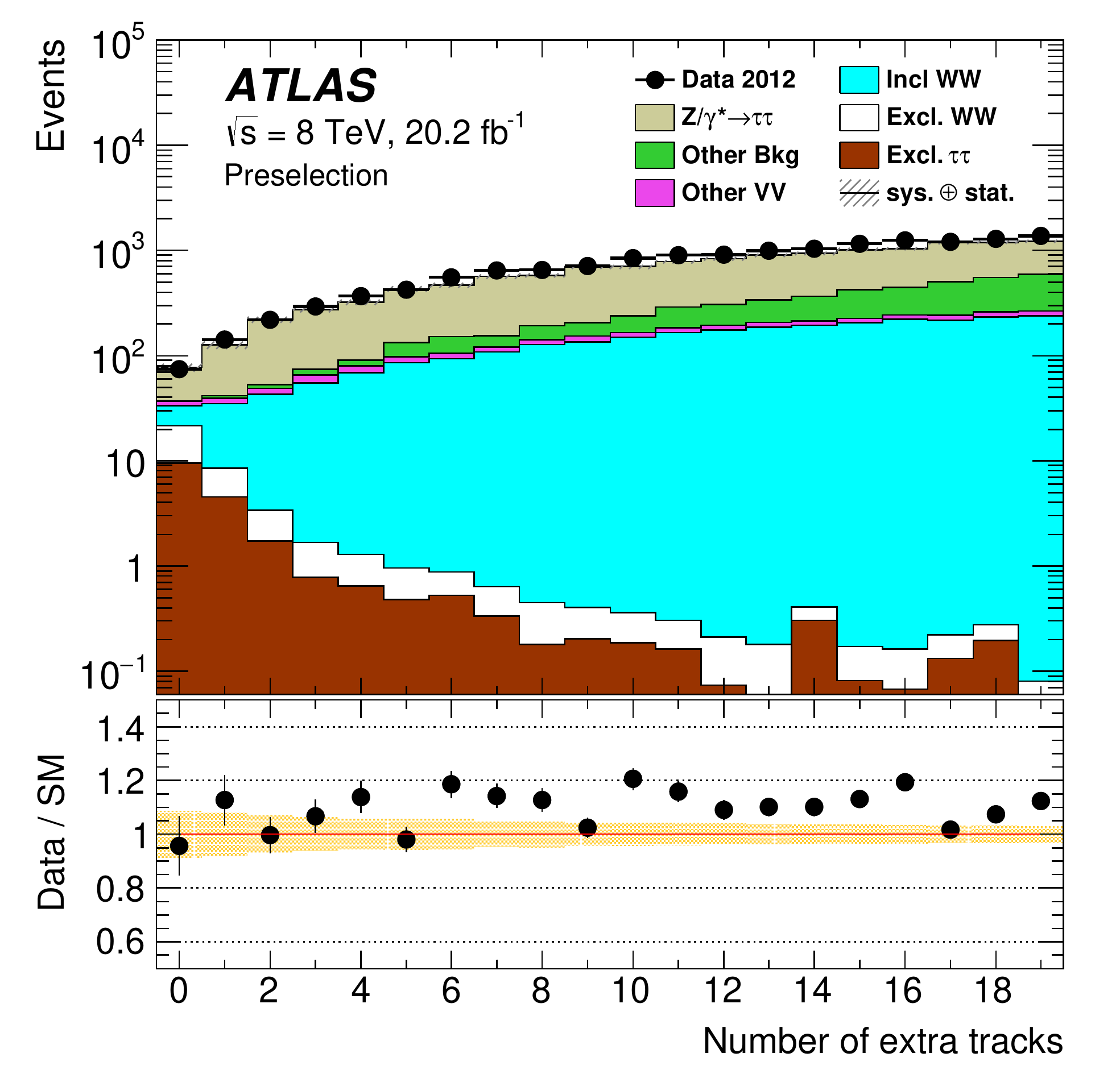} 
\end{minipage}
\hfill
\begin{minipage}{0.45 \linewidth}
 \includegraphics[width=\linewidth]{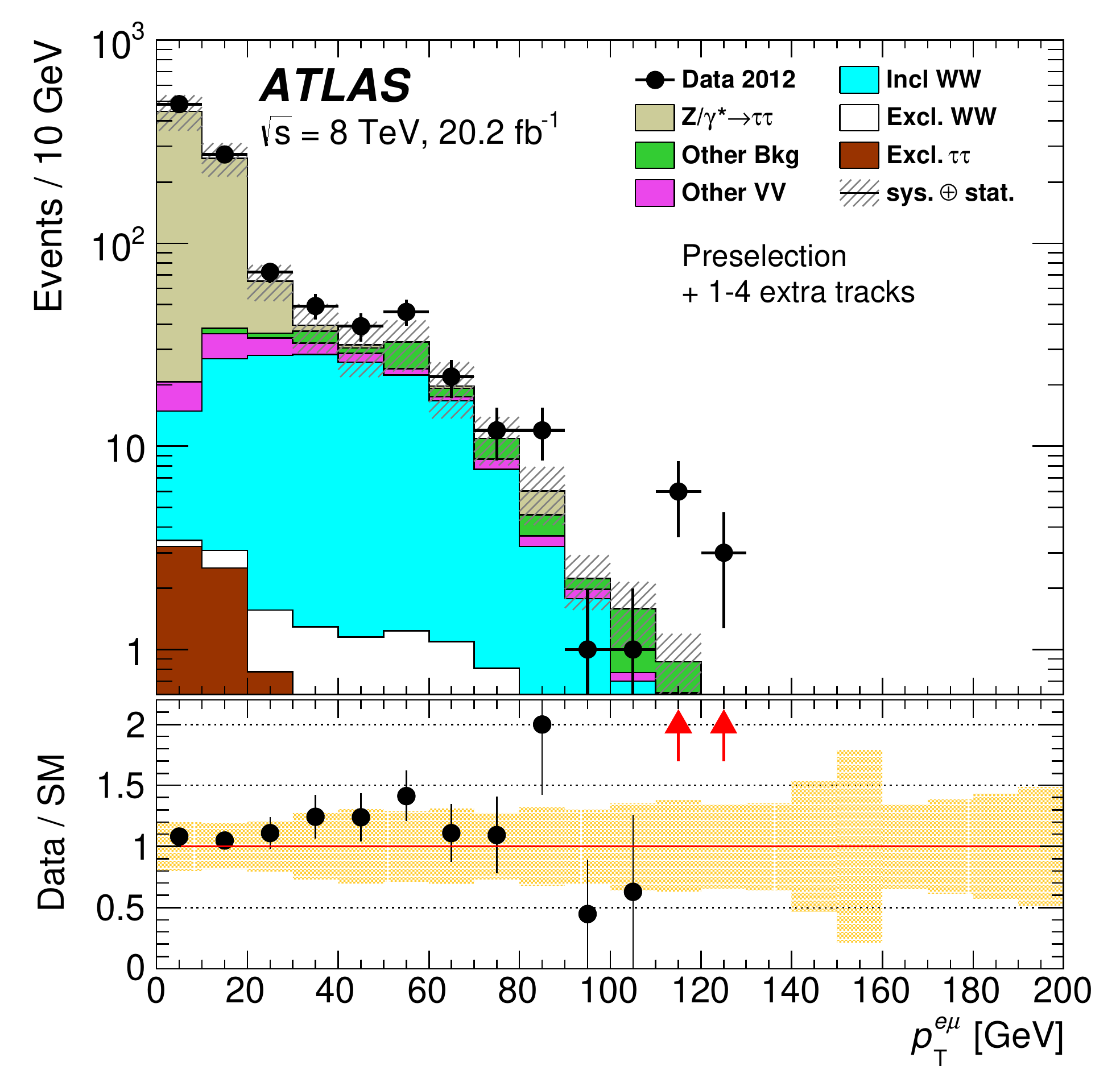} 
\end{minipage}
 \caption{Distribution of track multiplicities after requiring the
          exclusive $\WW$ preselection (left) with no number of track
          dependent correction,
          and the $\pT^{e\mu}$ distribution of candidates that have
          1--4 extra tracks (right), with
          the simulation including all appropriate correction factors such as $f^{\mathrm{sim}}_{\mathrm{nTracks}}$ (Table~\ref{tab:rejscale})
          for Drell-Yan and inclusive $\WW$
          production. 
          The enriched inclusive $\WW$ control region is the 1--4
          extra-track region above $\pT^{e\mu} > 30~\GeV$.
          The band around the Data/SM ratio of one illustrates the systemic uncertainties. 
          The upward red arrows indicate ratios outside the plotting range.}
 \label{fig:extraTracks} 
\end{figure}
\begin{figure}
\centering
 \includegraphics[width=0.5\linewidth]{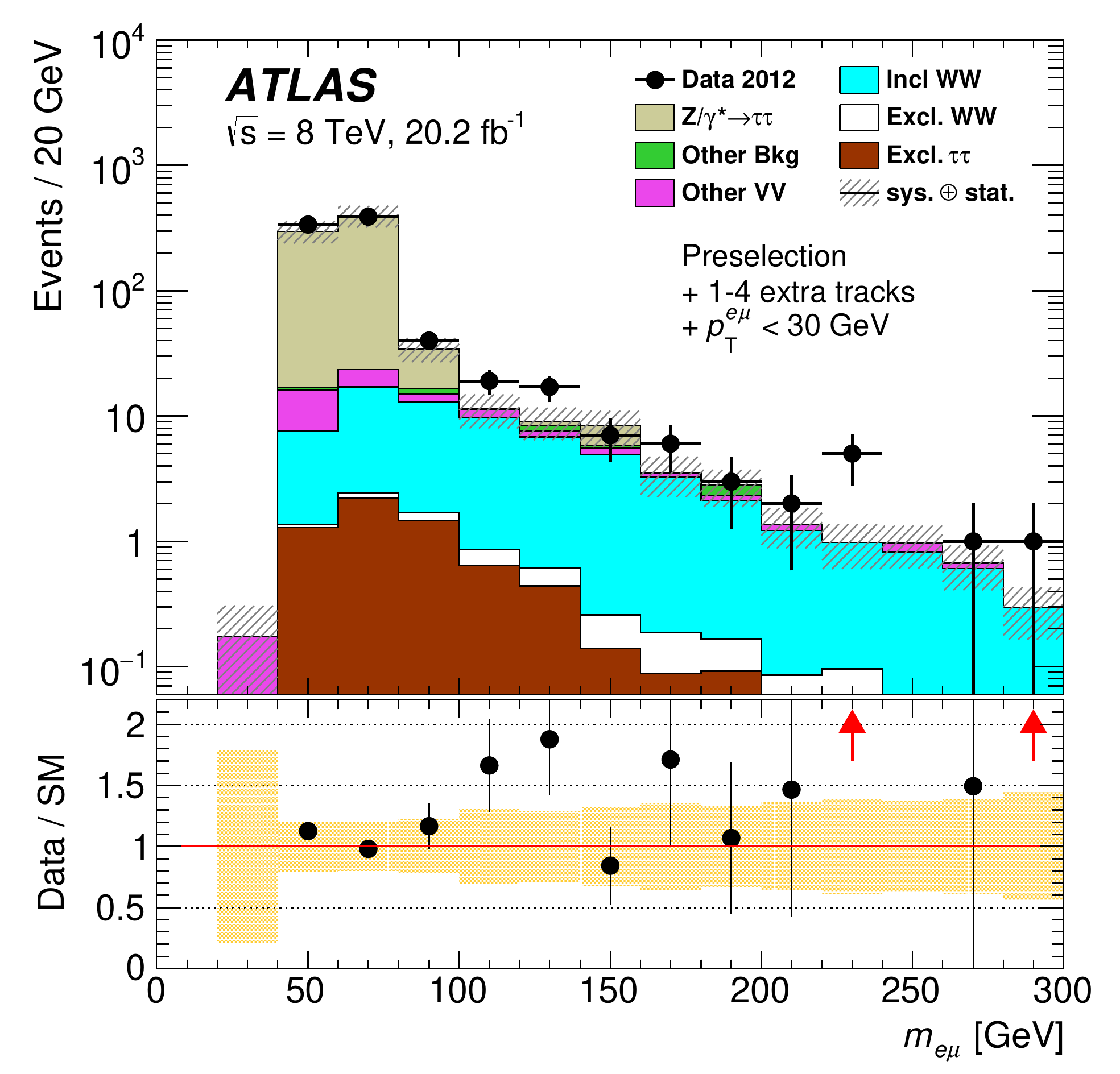} 
 \caption{The $m_{e\mu}$ distribution after requiring 1--4 extra tracks within 
          $\DZ = 1.0$~mm and $\pT^{e\mu} < 30~\GeV$. The Drell-Yan and inclusive $\WW$ samples 
          are scaled by the factors $\fAP_{1-4}$ and $\fP_{1-4}$, respectively. 
          The other samples are normalized as mentioned in the text.
          In the Data/SM ratio plot, the color band illustrates the systematic uncertainties, 
          the red upward arrows indicate ratios outside the plotting range.}
 \label{fig:xtrkZjetsMll} 
\end{figure}
\subsection{Inclusive \WW\ normalization}
  \label{sec:higgsControlRegions}
\par Inclusive \WW\ production is a significant background in both the 
exclusive Higgs boson and exclusive $\WW$ channels. 
From previous measurements~\cite{ATLAS:2014aga,Aad:2016wpd}, it it known that the NLO prediction for the $q\bar{q}\rightarrow W^{+}W^{-}$
 process as provided by \PowhegPythiaEight\ underestimates the observed $\WW$ event yield. 
It is therefore necessary to understand the simulation of this background before requiring the exclusivity selection. 
A region close in phase space to the exclusive Higgs boson signal region is chosen, referred to here as 
the Higgs-specific inclusive $\WW$ control region. It has the same definition except: $55 < \memu < 110~\GeV$,  
$\dFem<2.6$ to reduce Drell-Yan background, no jets to reduce \ttbar\ background, and no requirement on exclusivity.
This region is dominated by inclusive $\WW$ production and has a purity of 60\%. After subtracting the predicted backgrounds 
from data, $(20 \pm 5)\%$ more data is observed than is predicted
by \PowhegPythiaEight. A normalization factor of $1.20 \pm 0.05$(stat.)
is therefore taken as a correction to the cross-section and applied to the inclusive $\WW$ prediction in all regions 
of phase space studied here, as done in Ref.~\cite{ATLAS:2014aga}.
The transverse mass $\mT$ distributions in the Higgs-specific inclusive \WW\ control region
after applying the normalization factor to the \PowhegPythiaEight\ prediction is shown in Fig.~\ref{fig:higgsWWCRmT}.
\begin{figure}
 \centering
 \includegraphics[width=0.5\linewidth]{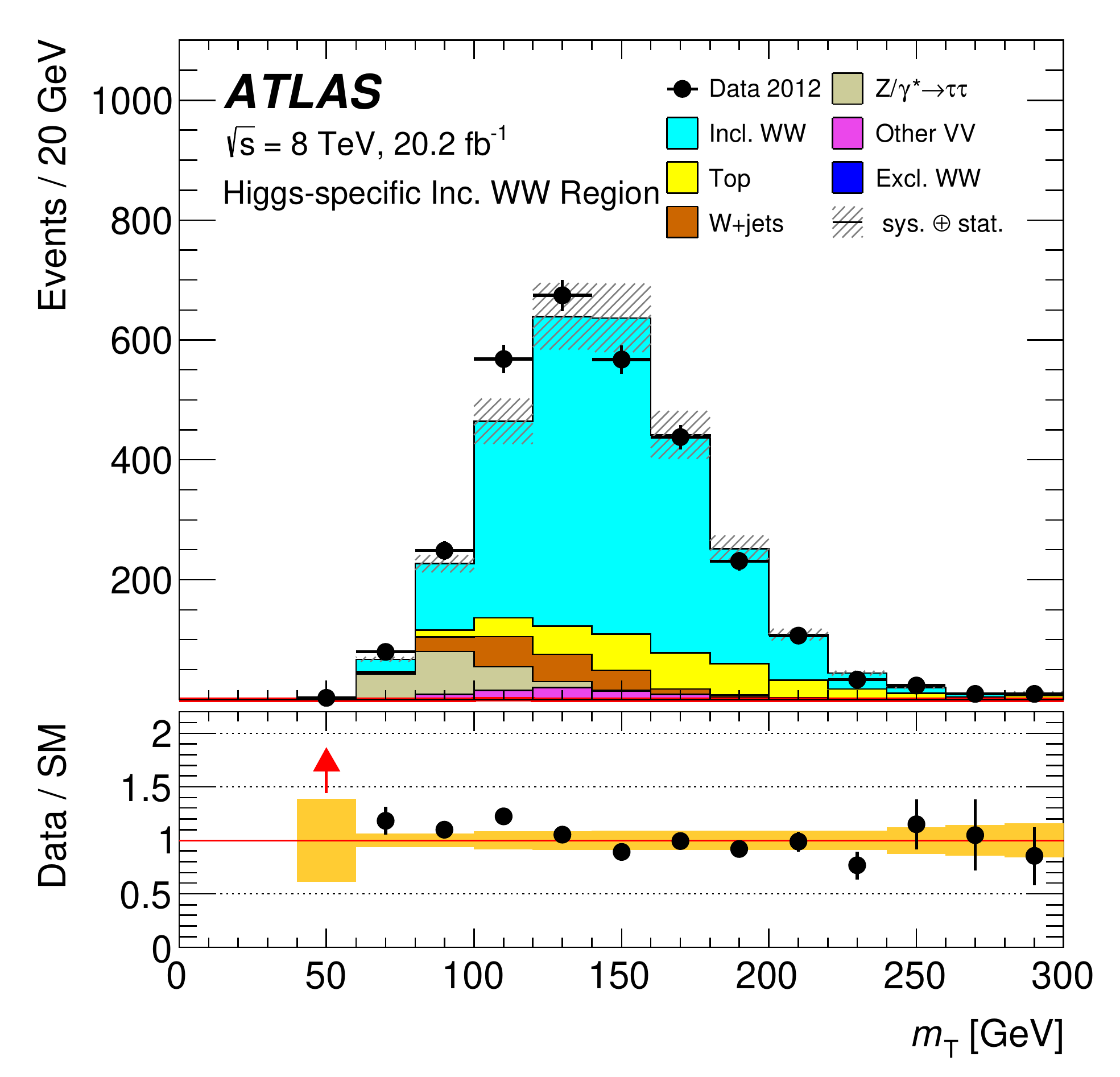} 
 \caption{The $\mT$ distributions in the Higgs-specific inclusive $W^+W^-$ control region 
		  that is used to determine the scaling for the \PowhegPythiaEight\ inclusive $\WW$ prediction.
          In the Data/SM ratio plot, the color band illustrates systematic uncertainties, 
          the red upward arrow indicates a ratio outside the plotting range.}
 \label{fig:higgsWWCRmT}
\end{figure}
\subsection{Sum of inclusive $\WW$ and other background}
  \label{sec:inclWWRegion}
An estimate of the sum of inclusive $\WW$ background and smaller contributions from Drell-Yan, 
$W$+jets, and top-quark production (collectively referred to as other background) is performed using 
an inclusive $\WW$-enriched control region defined with the same criteria as the exclusive $\WW$ 
signal region, except the exclusivity selection requires 1--4 extra tracks within 
$\DZ=1~\mm$. This control region is shown in Fig.~\ref{fig:extraTracks} (right) in the region above 
$\pT^{e\mu} > 30~\GeV$. It is dominated by the inclusive $\WW$ process and also has small 
contributions of exclusive events, non-$\WW$ (other-$VV$) dibosons, and other background.
\par Figure~\ref{fig:xtrkWWLeadingPt} shows the leading lepton $\pT^{\ell 1}$ distribution in this control region. The prediction 
is systematically lower than the data. The processes contributing to this control region can be 
found in Table~\ref{tab:xtrkWWEventYields}, and the total SM
expectation is compared to the data. The data exceeds the simulation by $2\sigma$. This discrepancy is attributed 
to a component from jets faking leptons that is unreliably simulated. Events produced with jets such 
as $W$+jets, $Z$+jets, and top-quark production, particularly jets faking leptons, are more easily rejected 
by the exclusivity selection, while other-$VV$ and Drell-Yan (without accompanying jets) processes 
are likely to extrapolate from the 1--4 extra-track control region to the zero-track region with  
a scale factor similar to that for inclusive $\WW$ background. Therefore, this control region is used to constrain 
the inclusive $\WW$ plus other background involving fake leptons. 
\begin{figure}
\centering
 \includegraphics[width=0.5\textwidth]{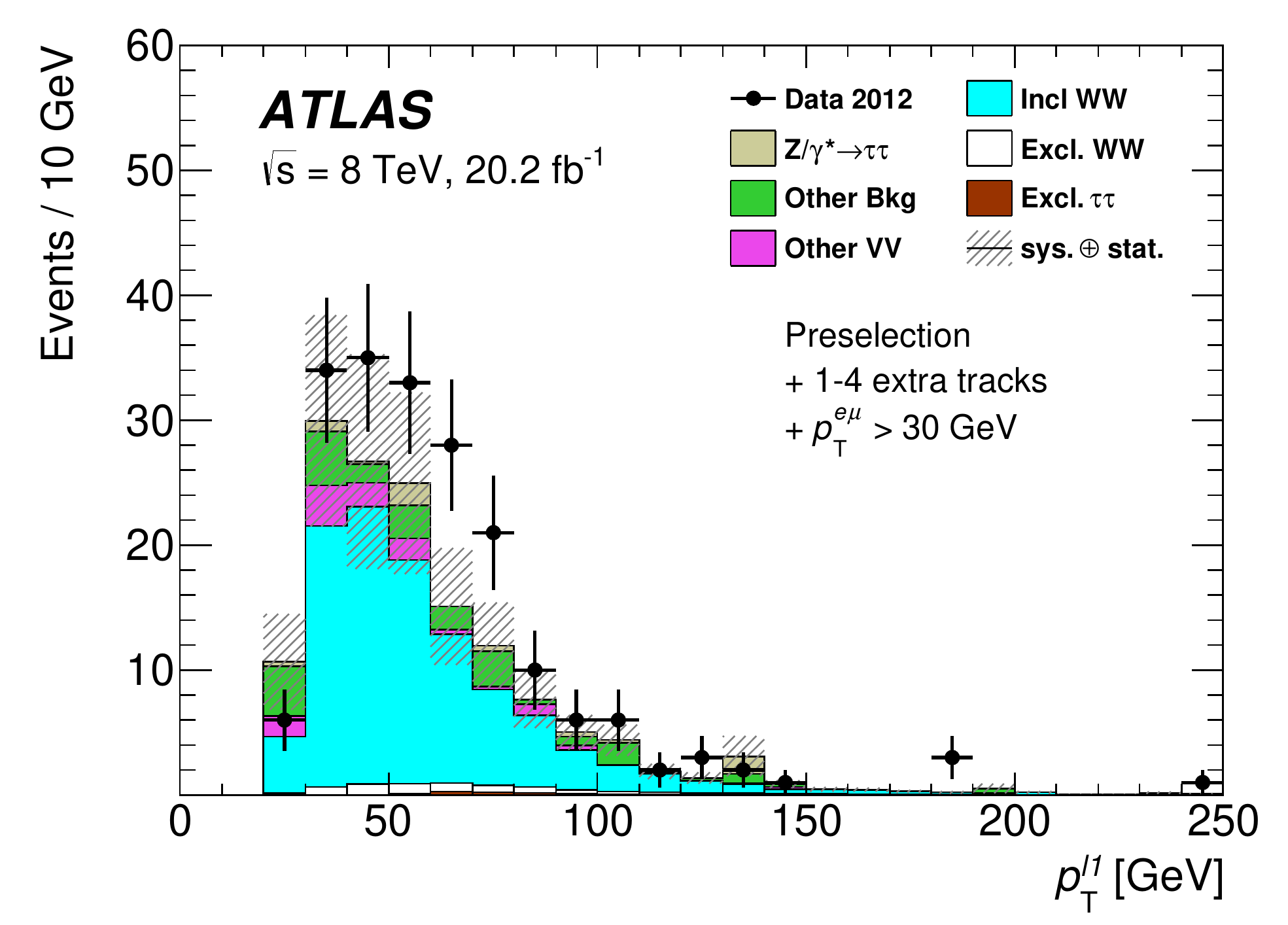} 
 \caption{The leading-lepton $\pT^{\ell 1}$ distribution in the inclusive $\WW$ control region. 
 The simulation includes all appropriate correction factors such as $f^{\mathrm{sim}}_{1-4}$ for Drell-Yan and $\fP_{1-4}$ 
 for inclusive $\WW$ production.}
 \label{fig:xtrkWWLeadingPt} 
\end{figure}
\begin{table}[!h]
   \centering
   \begin{tabular}{l|c}
     \hline\hline
    Processes & Inclusive $\WW$ \\
    \hline\hline
    Inclusive $\WW$             & 102 $\pm$ 20 \\
    Exclusive $\WW$             & 5.5 $\pm$ 0.4 \\
    Exclusive $\tau^+\tau^-$    & 1.2 $\pm$ 0.2 \\
    Other diboson               & 10.9 $\pm$ 2.2  \\
    Other background            & 27.4 $\pm$ 6.2 \\
    \hline
    Total SM                    & 147 $\pm$ 21  \\
    Data                        & 191             \\
    \hline\hline                                  
   \end{tabular}
  \caption{Event yields in the inclusive $\WW$ control region.
           The uncertainties quoted are statistical and systematic.}
  \label{tab:xtrkWWEventYields}
\end{table}
\par For the purpose of estimating the contribution of inclusive $\WW$ events and other background in the zero-track region, 
the number of these events in the 1--4 extra-track control region is bracketed by the number of
observed events in the data, after subtracting the exclusive and other-$VV$ contributions, as an 
upper bound and by the predicted number of inclusive $\WW$ obtained from $\PowhegPythiaEight$ as a lower 
bound. To obtain the contribution for the exclusive $\WW$ signal region, the two estimates are 
extrapolated from the 1--4 extra-track control region to the zero-track signal region. In this 
framework, the lower bound corresponds to the optimistic case where the other background 
contribution is completely rejected by the zero-track exclusivity requirement, while the upper 
bound corresponds to the case where all observed candidates in the control region are suppressed 
by the same factor as the inclusive $\WW$ process. Finally the average of the two estimates (after 
extrapolation) is taken as the contribution for the signal region.
\par The extrapolation is achieved by multiplying the estimates by the ratio of the predicted numbers 
of inclusive $\WW$ events:
\begin{equation}
N^{\textrm{Estimated}}_{0} = N^{\textrm{Estimated}}_{1-4} 
                       \times \frac{N_{WW,0}^{\textrm{Predicted}}}{N_{WW,1-4}^{\textrm{Predicted}}},
\end{equation} 
where $N^{\textrm{Estimated}}_{0}$ and $N^{\textrm{Estimated}}_{1-4}$ 
are the estimates for the lower bound or upper bound mentioned above, and $N_{WW,0}^{\textrm{Predicted}}$ 
and $N_{WW,1-4}^{\textrm{Predicted}}$ are respectively the number of inclusive $\WW$ events predicted 
by $\PowhegPythiaEight$ for the zero-track and 1--4 extra-track regions. 
This ratio is found to be 
$0.048 \pm 0.014$, where the uncertainty is dominated by the 20\% systematic uncertainties taken 
to be uncorrelated between the $\fP_{0}$ and $\fP_{1-4}$ factors that are included in the 
predicted numbers of events. As mentioned above, the small exclusive and other-$VV$ contributions are 
subtracted before the extrapolation. So for inclusive $\WW$ and Drell-Yan processes, 
the expected number of events in the zero-track region is 20 times less than the prediction for 
the 1--4 extra-track control region.
\par As mentioned above, the inclusive $\WW$ and other background contributions to the signal region are 
taken as the average of the two estimates. Half the difference is included as an additional 
contribution to the uncertainty in this determination. This results in a final estimate of 
$6.6 \pm 2.5$ background candidates for the exclusive $\WW$ signal region.
\par This background estimate, $6.6 \pm 2.5$ events in the exclusive $\WW$ signal region, corresponds 
to scaling the $\PowhegPythiaEight$ $\WW$ prediction by a normalization factor of 0.79. 
This factor is used to estimate the inclusive $\WW$ and other background contamination in 
the Higgs boson and aQGC signal regions.

\section{Systematic uncertainties}
\label{sec:systs}
\par The main sources of systematic uncertainty are related to the exclusivity selection and the
background determination. The uncertainty in the efficiency of the
exclusive signal selection contributes 10\% to the 
exclusive $W^+W^-$ and Higgs boson signal yields, as estimated in Section~\ref{sec:validationDilepton}
from the ratios of dimuon event yields without extra tracks and those with exactly one extra track.
The prediction of the exclusive $\WW$ process uses the
$\fgam$ factor as described in Section~\ref{sec:photonFlux} and thus
carries the $7\%$ uncertainty in $\fgam$.  
The $\yytautau$ background has an uncertainty of 14\% that is propagated from the $\fEL$ factor. 
As described in Section~\ref{sec:validationDilepton}, the $\fEL$ uncertainty includes 10\% related to
the exclusive signal selection and another 10\% that results from acoplanarity fits. 
There is a 38\% uncertainty in the 
inclusive $\WW$ background, as discussed in Section~\ref{sec:inclWWRegion}.
This 38\% uncertainty contains a component from the $\pm 20\%$ 
uncertainty in Drell-Yan background described in Section~\ref{sec:MCTrackModelling}.
\par The contributions from these systematic uncertainties to the measured exclusive $\WW$ 
cross-section can be found in Table~\ref{tab:mainSystematics}. The overall background contribution 
is 18\%, predominantly from uncertainty in the extrapolation from the
1--4 track
control region. In addition to the systematic uncertainty from the exclusivity selection (10\%), 
other systematic uncertainties (lepton selection efficiencies and acceptance, luminosity and 
lepton scales and resolution) contribute less than 5\%. 
The statistical uncertainty dominates the uncertainties in the cross-section. 
\begin{table}
\centering
\begin{tabular}{l|c}
\hline
\hline
Source of uncertainty & Uncertainty [$\%$]\\
\hline
Statistics & $33\%$ \\
Background determination & $18\%$ \\ 
Exclusivity signal efficiency & $10\%$ \\
All other & $<5\%$ \\
\hline
Total & $39\%$\\
\hline 
\hline
\end{tabular}
\caption{Sources of uncertainty for the measured exclusive $\WW$ cross-section.
	All other includes other
    efficiencies, acceptance, luminosity, and lepton scales and
    resolution. 
	}
\label{tab:mainSystematics}
\end{table}
\begin{figure}[!h]
 \centering
  \includegraphics[width=0.45\linewidth]{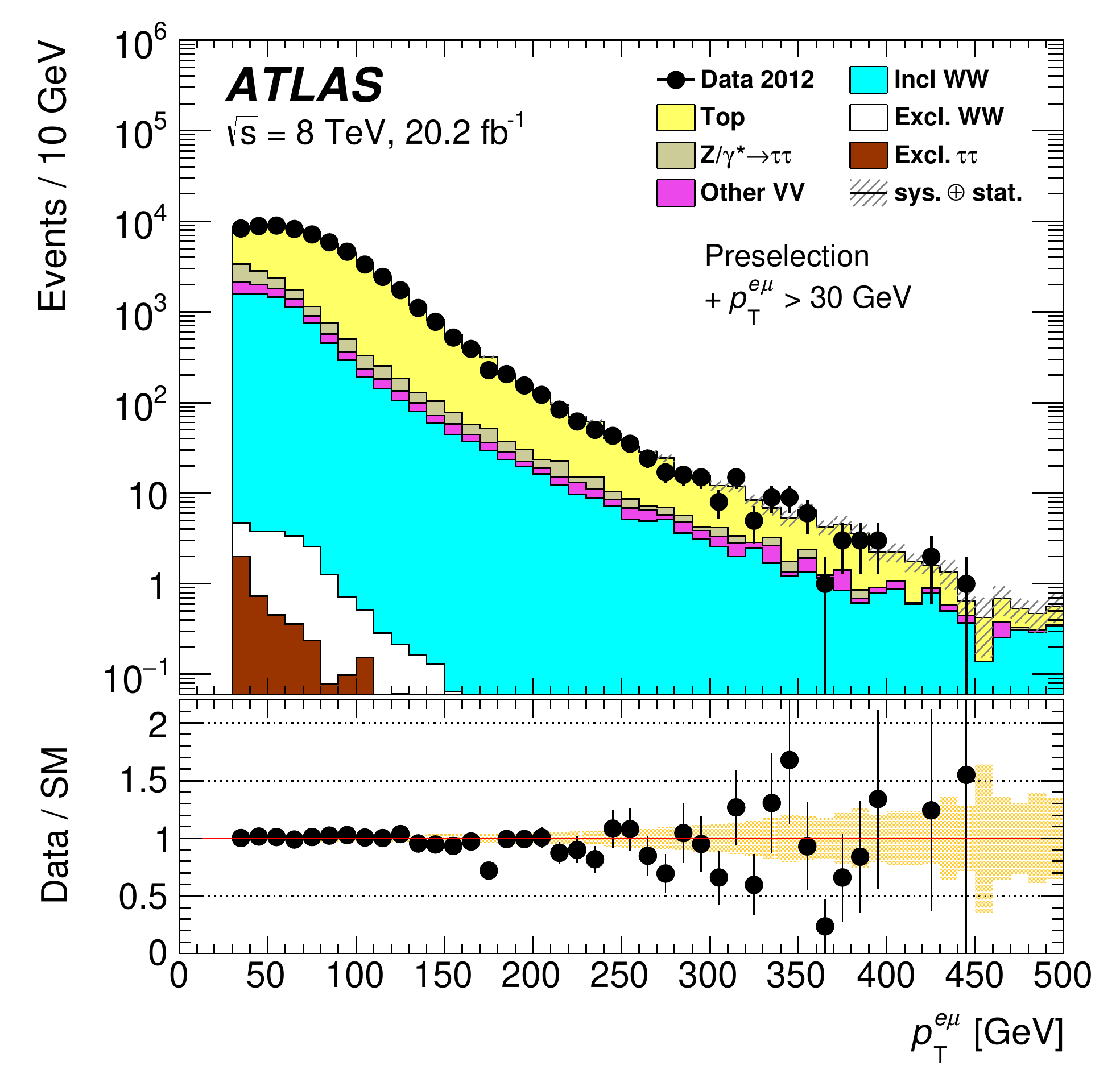}
 \caption{The $p_{\mathrm{T}}^{e\mu}$ distribution before exclusivity, i.e.,
   after requiring $p_{\mathrm{T}}^{e\mu} > 30$~\GeV. The main
   backgrounds at this stage are top-quark production, inclusive $\WW$ and
   Drell-Yan.
    In the Data/SM ratio plot, the color band illustrates systematic uncertainties.}
 \label{fig:exclWWkinematicBeforeExcl}
\end{figure}

\section{Results}
\label{sec:results}
This paper presents three main results: the exclusive $\WW$ production cross-section,
limits on possible aQGCs, and a limit from a search for exclusive Higgs boson
production. Each is summarized in the following. 
The exclusive $\WW$ signal is the sum of elastic and single- and double-dissociative 
events through the $\fgam$ factor discussed in Section~\ref{sec:photonFlux}.
\subsection{Standard Model exclusive $\WW$ production}
\label{sec:SMResults}
\par Before the exclusivity selection,
good agreement between data and background prediction is 
observed. In the $e\mu$ final state, the overall event yield agrees to
within 2\% and after requiring $p_{\mathrm{T}}^{e\mu}> 30$~\GeV it
agrees to within $0.5\%$. 
The $\pTemu$ distribution before the
exclusivity requirement is shown in Fig.~\ref{fig:exclWWkinematicBeforeExcl}.  
\par The numbers of candidates at various stages of the analysis are listed in
Table~\ref{tab:exclWWCutflow} and the uncertainties quoted include
both the statistical and systematic uncertainties. 
Top-quark and Drell-Yan $\DrellYantotautau$ processes are the dominant backgrounds before exclusivity, while
after requiring exclusivity their contributions are less than 0.5 events. 
These two backgrounds along with $W$+jets are grouped together as other background 
(Table~\ref{tab:exclWWCutflow}).
The inclusive $\WW$ estimate (described in Section~\ref{sec:inclWWRegion}) 
already includes these three processes; thus the other background contribution after requiring 
exclusivity is not added to the total background. 
Non-$\WW$ (other-$VV$) diboson 
processes are also highly suppressed by the exclusivity selection: they contribute $0.3 \pm 0.2$ events. 
Diffractive $\WW$ production was considered as a background and found to be
insignificant. 
The expected signal yield is $9.3 \pm 1.2$ events, including the dissociative contributions 
($\fgam$ factor) discussed in Section~\ref{sec:photonFlux}.
The total predicted background is $8.3 \pm 2.6$, while 23 candidates are observed in the data. 
\par Figure~\ref{fig:exclWWkinematicNominal} shows the $\pTemu$ and $\Delta\phi_{e\mu}$
distributions after applying all selection criteria. The shapes of the signal
and the inclusive $\WW$ distributions are similar. 
The remaining $\tau^+\tau^-$ background has an azimuthal
opening angle close to $\Delta \phi_{e\mu} \sim \pi$, i.e., the leptons 
are back-to-back. 
No further requirement is applied to $\Delta \phi_{e\mu}$ to reject this background, 
as the aQGC signal also has an enhancement for $\Delta \phi_{e\mu} \sim \pi$.
\begin{table}[!h]
\begin{center}
\resizebox{\textwidth}{!}{
\begin{tabular}{l||rrr|rrrr||rr}
\hline\hline
 & Expected Signal & Data & Total Bkg  & Incl $\WW$ & Excl. $\tau\tau$ & Other-$VV$ &
Other Bkg & SM/Data &  $\epsilon A$ (Signal)\\
\hline\hline
 Preselection                        & $22.6 \pm 1.9$ & 99424    & 97877         & 11443         & 21.4            & 1385             & 85029    & 0.98  & 0.254 \\
 $p_{\mathrm{T}}^{\ell\ell} > 30$~\GeV& $17.6 \pm 1.5$ & 63329    & 63023         & 8072          & 4.30            & 896.3            & 54051    & 1.00  & 0.198 \\
 $\DZ$ requirement                   & $9.3 \pm 1.2$  & 23       & $8.3 \pm 2.6$ & $6.6 \pm 2.5$ & $1.4 \pm 0.3$ & $0.3 \pm 0.2$  & $-$    & 0.77  & $0.105 \pm 0.012$ \\
 \hline
 aQGC signal region \\
 \hline
 $p_{\mathrm{T}}^{\ell\ell} > 120$~\GeV & $0.37 \pm 0.04$ & 1     & $0.37 \pm 0.13$       & $0.32 \pm 0.12$  & $0.05 \pm 0.03$ & 0 & $-$ & 0.74 & $0.0042 \pm 0.0005$ \\
\hline\hline
\end{tabular}
}
\end{center}
\caption{The event yield at different stages of the selection. The
  expected signal ($\yyWW$) is compared to the data and total background. 
  The SM-to-data ratio (SM/Data) gives the level of agreement between prediction 
  and data. The product of efficiency and acceptance ($\epsilon A$) for 
  the signal is computed from the $\yyWW \to \emu$ MC generator.
  The statistical and systematic uncertainties are 
  added in quadrature. For the background, the uncertainties are only
  shown for the yields 
  after exclusivity selection, where they are relevant for the measurement.}
\label{tab:exclWWCutflow}
\end{table}
\begin{figure}
 \centering
 \begin{tabular}{cc}
  \includegraphics[width=0.5\linewidth]{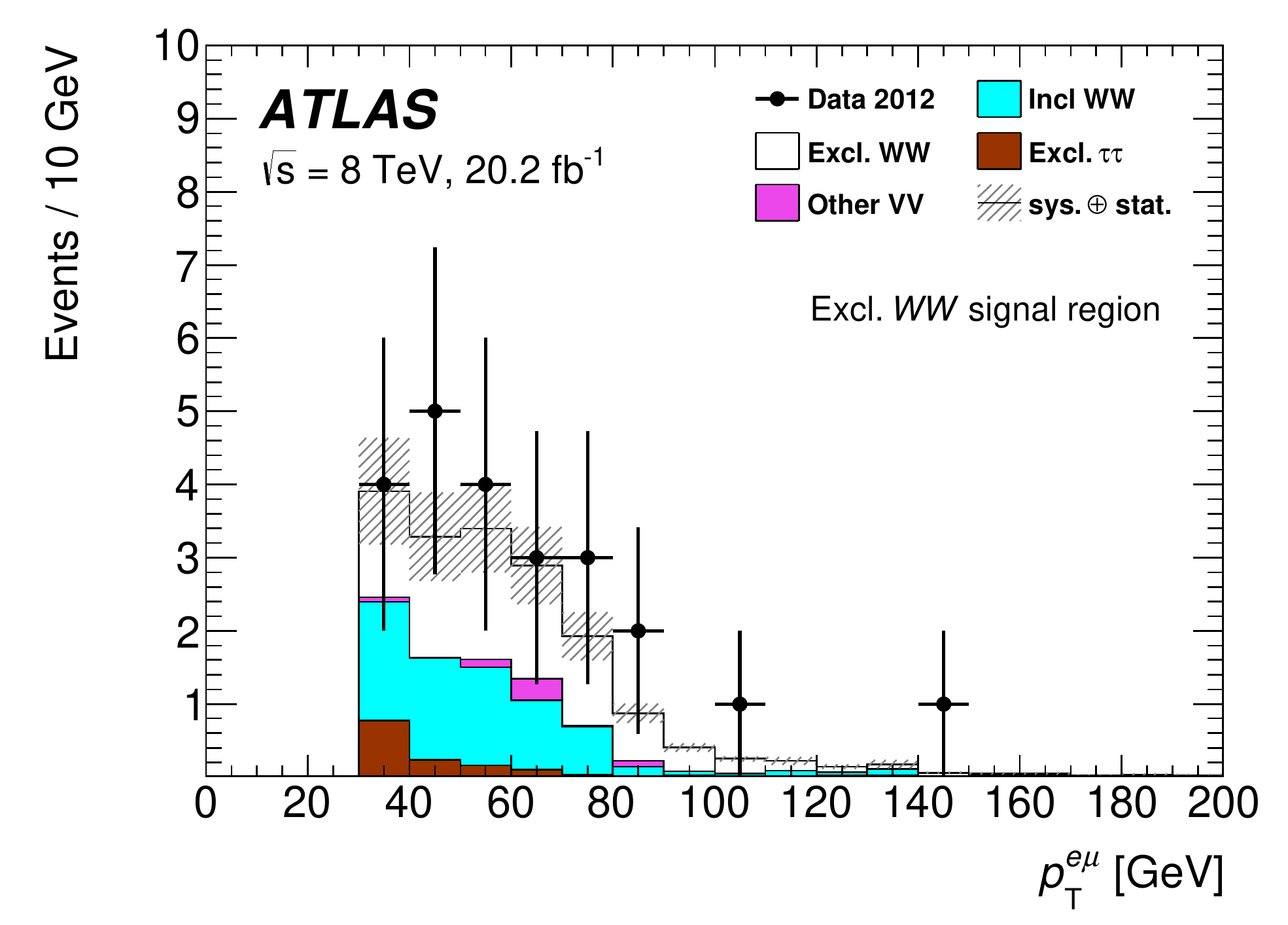}
  \includegraphics[width=0.5\linewidth]{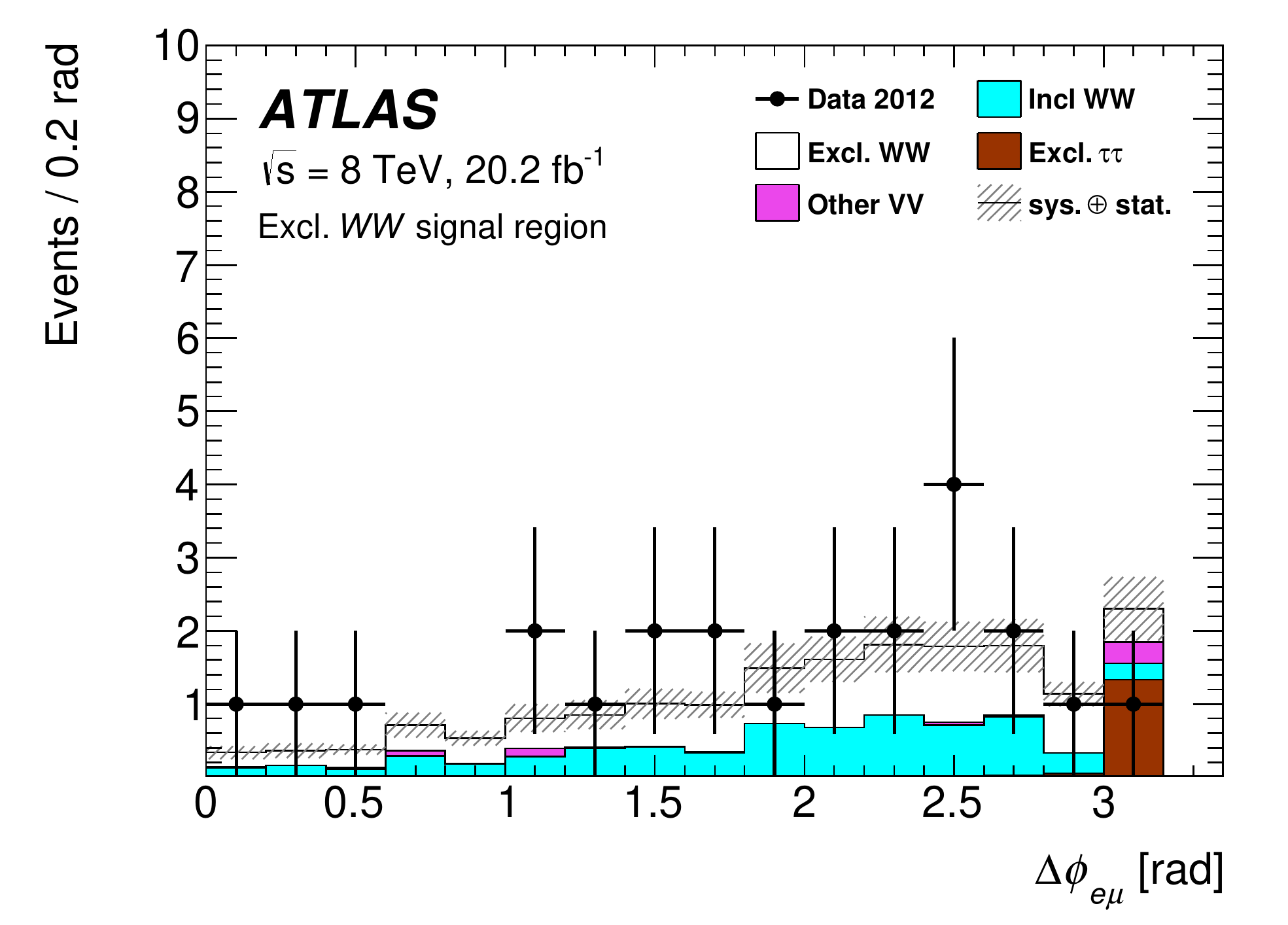}
 \end{tabular}
 \caption{The \pTemu (left) and $\Delta\phi_{e\mu}$ (right) distributions in the exclusive $\WW$ signal region.
    The inclusive $\WW$ estimate includes small contributions from other backgrounds (Drell-Yan, $W$+jets, and top-quark production).
}
 \label{fig:exclWWkinematicNominal}
\end{figure}
\subsubsection{$\yyWW$ cross-section}
\label{sec:exclWWCrossSection}
\par The full phase space cross-section predicted by \HERWIGPP is
$\sigma_{\yyWW}^{\HERWIGPP} = 41.6$~fb. This number is well defined
but $\sim$20\% corrections similar to those for the EPA dilepton prediction are expected, 
as discussed with Equation~(\ref{eqn:fELfactor}) above. 
The branching ratio of the $\WW$ pair decaying to $e^{\pm}\mu^{\mp}X$ 
is $\BR(\WW \to e^{\pm}\mu^{\mp} X) = 3.23\%$~\cite{PDG} (including the
leptonic decays of $\tau$ leptons). Therefore, the predicted cross-section corrected for 
$\BR(\WW \to e^{\pm}\mu^{\mp} X)$ and including the 
dissociative contributions through the normalization $\fgam = 3.30 \pm 0.23$ becomes:
\begin{equation}
 \sigma_{\yyWW \rightarrow e^{\pm}\mu^{\mp} X}^{\textrm{Predicted}} 
        = \fgam \cdot  \sigma_{\yyWW}^{\textrm{\HERWIGPP}} \cdot \BR(\WW \to e^{\pm}\mu^{\mp} X) 
        = 4.4 \pm 0.3~\textrm{fb},
 \label{eqn:predictedXS}
\end{equation}
which corresponds to the prediction of $N_{\textrm{Predicted}} = 9.3 \pm
1.2$ signal events, quoted in Table~\ref{tab:exclWWCutflow}.
The number of candidates observed in the data is $N_{\textrm{Data}}= 23$,
while the predicted background is $N_{\textrm{Background}} = 8.3 \pm 2.6$ events. 
So the observation exceeds the prediction by a ratio:
\begin{equation}
 R = (N_{\textrm{Data}} - N_{\textrm{Background}})/N_{\textrm{Predicted}} 
   = 1.57 \pm 0.62.
\label{eqn:ratioR}
\end{equation}
The uncertainty in $R$ 
results from propagation of the uncertainties of each of the numbers that go into the calculation.
The uncertainty in the factor $\fgam$ contributes 7\%. 
\par The measured cross-section is 
determined in the exclusive $\WW$ region and extrapolated to 
the full $\WW \to e^{\pm} \mu^{\mp} + X$ phase space: 
\begin{equation}
 \sigma_{\yyWW \rightarrow e^{\pm}\mu^{\mp} X}^{\textrm{Measured}}
   = (N_{\textrm{Data}} - N_{\textrm{Background}})/ (\mathcal{L}\; \epsilon A)
   = 6.9 \pm 2.2~\textrm{(stat.)} \pm 1.4~\textrm{(sys.)}\ \textrm{fb},
\label{eqn:measuredXS}
\end{equation}
where $\mathcal{L} = 20.2 \pm 0.4~\ifb$. 
The acceptance ($A$) is the ratio of the number of simulated events passing 
the kinematic requirements in Table~\ref{tab:selections} to the total number of events generated. The efficiencies ($\epsilon$) 
account for the detector efficiencies due to lepton identification and reconstruction, 
triggering, and pileup.
Both $A$ and $\epsilon$ are computed using the $\HERWIGPP$                                                         
prediction for the elastic $\yyWW$ process.                                   
At the end of the event selection, the acceptance is $A = 0.280 \pm 0.001$ 
and the efficiency, which includes the exclusivity selection efficiency, 
is $\epsilon = 0.37 \pm 0.04$.
\par The efficiency of the exclusivity selection is $0.58 \pm 0.06$.                 
The elastic, SD, and DD predicted acceptances can be compared using
$\yymumu$ events                                             
with $m_{\mu\mu} > 160~\GeV$, and they are found to be the same within     
$3\%$. Therefore, the measurement of the cross-section can be
performed with the 
acceptances for elastic $\yyWW$ events.                                           
The products of acceptance and efficiencies ($\epsilon A$) at
different           
stages of the event selection are given in
Table~\ref{tab:exclWWCutflow}.      
\par The sources of uncertainty are given in  
Table~\ref{tab:mainSystematics}. The statistical uncertainty
dominates. The contribution from intermediate $\tau$ leptons to the accepted signal MC
is determined using the $\HERWIGPP$ generator to be $9.1\%$. 
The background-only hypothesis has a $p$-value about 0.0012, 
corresponding to a significance of 3.0$\sigma$.
\subsubsection{Limits on anomalous Quartic Gauge Couplings}
\par The aQGC limit setting was performed using the region $\pTemu > 120~\GeV$ where the aQGC contributions are expected
to be important and Standard Model backgrounds are suppressed.
The $\pTemu$ distribution is shown in Fig.~\ref{fig:exclWW_pTllAQGC}
for data compared to the Standard Model prediction and various aQGC
scenarios.  The aQGCs enhance the exclusive signal at high $\pTemu$, 
while the background is negligible with $\pTemu > 80~\GeV$. 
The 95\% CL limits on the couplings $a_0^W/\Lambda^2$ and $a_C^W/\Lambda^2$ are extracted 
with a likelihood test using the one observed data event as a constraint. 
 \begin{figure}
  \centering
  \begin{tabular}{cc}
   \includegraphics[width=0.5\textwidth]{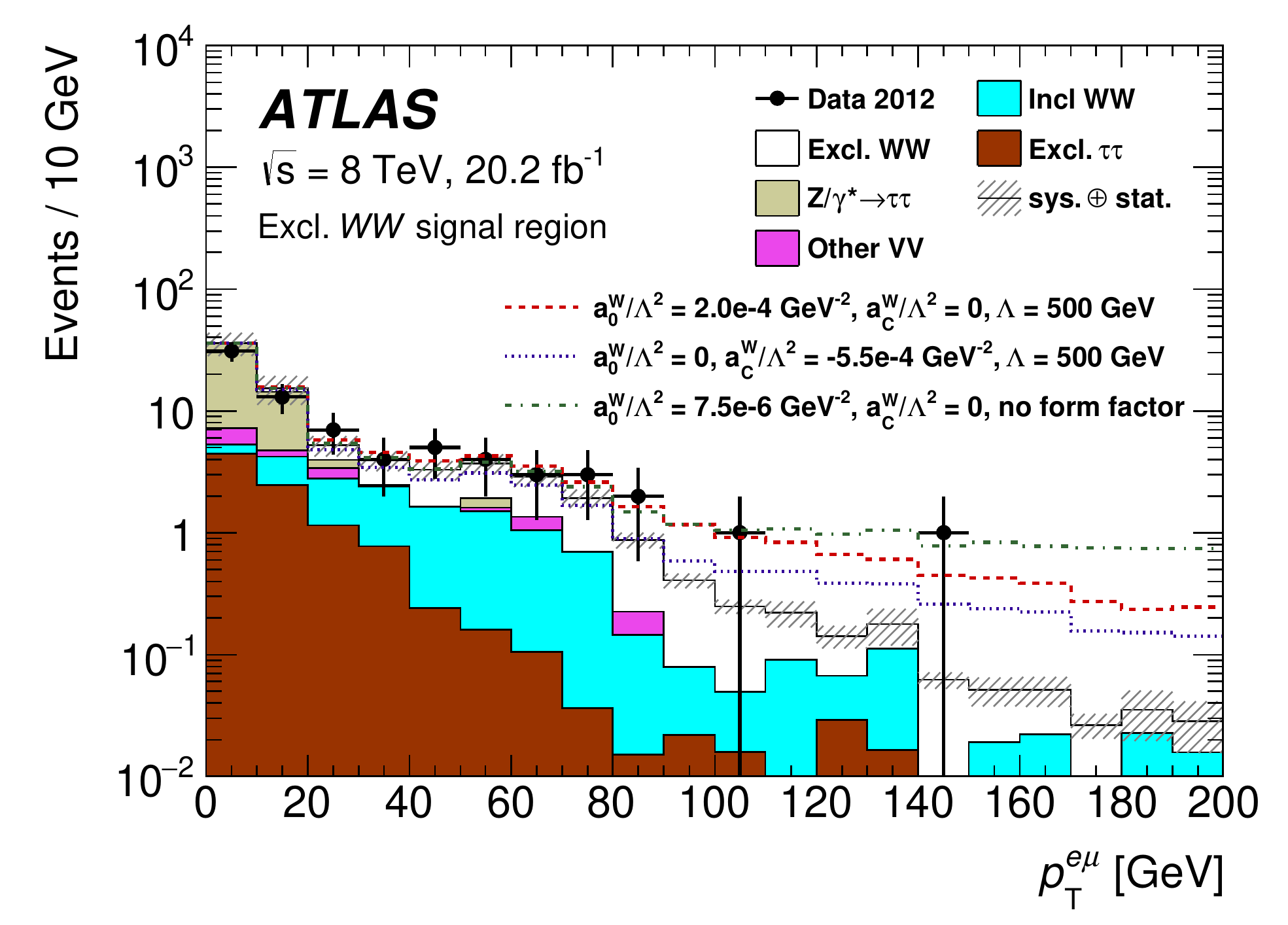} 
  \end{tabular}
  \caption{The $\pT^{e\mu}$ distribution for data compared to the SM prediction for events satisfying all the 
           exclusive $\WW$ selection requirements apart from the one on $\pT^{e\mu}$ itself. Also
           shown are various predictions for aQGC parameters $a_{0,C}^W$. 
           }
  \label{fig:exclWW_pTllAQGC}
 \end{figure}
\par To extract one-dimensional (1D) limits, one of the aQGCs is set to zero. 
The 95\% CL allowed ranges for the cases with a dipole form factor defined in Equation~(\ref{eqn:dipoleff}) 
with $\LambdaCutoff = 500~\GeV$ and without a form factor ($\LambdaCutoff \to \infty$) 
are listed in Table~\ref{tab:aQGClimits}.
The uncertainties 
in the yields are included in the likelihood test as nuisance parameters.
Also, limits on the two aQGC parameters are shown in Fig.~\ref{fig:aQGCcontour} for the case 
with a dipole form factor with $\LambdaCutoff = 500~\GeV$. The region outside the contour  
is ruled out at 95\% confidence-level. The limits are comparable to
the CMS combined 7 and 8~\TeV results~\cite{CMSyyWW7and8}.
\begin{table}
 \centering
  \renewcommand*{\arraystretch}{1.1} 
  \begin{tabular}{lccc}
   \hline\hline
    Coupling & $\LambdaCutoff$ & Observed allowed range [\GeV$^{-2}$] &
    Expected allowed range [\GeV$^{-2}$]\\
    \hline
    ${a_0^W}/{\Lambda^2}$ & $500~\GeV$ & $[-0.96\times
      10^{-4},\ 0.93\times 10^{-4}]$ & $[-0.90\times
      10^{-4},\ 0.87\times 10^{-4}]$\\
    ${a_C^W}/{\Lambda^2}$ & $500~\GeV$ & $[-3.5\times
      10^{-4},\ 3.3\times 10^{-4}]$  & $[-3.3 \times 10^{-4},\ 3.1
      \times 10^{-4}]$ \\
    \hline
    ${a_0^W}/{\Lambda^2}$ & $\infty$ & $[-1.7\times 10^{-6},\ 1.7\times 10^{-6}]$ & $[-1.5\times 10^{-6},\ 1.6\times 10^{-6}]$\\
    ${a_C^W}/{\Lambda^2}$ & $\infty$ & $[-6.4\times 10^{-6},\ 6.3\times 10^{-6}]$& $[-5.9\times 10^{-6},\ 5.8\times 10^{-6}]$\\
    \hline\hline
  \end{tabular}
 \caption{The observed allowed ranges for ${a_0^W}/{\Lambda^2}$ and ${a_C^W}/{\Lambda^2}$,
  for dipole form factor with $\LambdaCutoff = 500$~\GeV and without form factor ($\LambdaCutoff \to \infty$). 
  The regions outside the quoted ranges are excluded at 95\% confidence-level.}
 \label{tab:aQGClimits}
\end{table}
\begin{table}
 \centering
  \renewcommand*{\arraystretch}{1.1} 
  \begin{tabular}{lccc}
   \hline\hline
    Coupling & $\LambdaCutoff$ & Observed allowed range [$\GeV^{-4}$] & Expected allowed range [$\GeV^{-4}$] \\
    \hline
    $f_{M,0}/{\Lambda^4}$ & $500~\GeV$ & $[-3.7\times 10^{-9},\ 3.6 \times 10^{-9}]$ & $[-3.5\times 10^{-9},\ 3.4 \times 10^{-9}]$ \\
    $f_{M,1}/{\Lambda^4}$ & $500~\GeV$ & $[-13 \times 10^{-9},\ 14  \times 10^{-9}]$ & $[-12 \times 10^{-9},\ 13  \times 10^{-9}]$\\
    \hline
    $f_{M,0}/{\Lambda^4}$ & $\infty$ & $[-6.6\times 10^{-11},\ 6.6 \times 10^{-11}]$ & $[-5.8\times 10^{-11},\ 6.2 \times 10^{-11}]$ \\
    $f_{M,1}/{\Lambda^4}$ & $\infty$ & $[-24 \times 10^{-11},\ 25  \times 10^{-11}]$ & $[-23 \times 10^{-11},\ 23  \times 10^{-11}]$ \\
    \hline\hline
  \end{tabular}
 \caption{The allowed ranges for dimension-8 couplings values derived from the $a_0^W$ and $a_C^W$
          parameters, for a dipole form factor with $\LambdaCutoff = 500$~\GeV and without form factor. 
          The regions outside the quoted ranges are excluded at 95\% confidence-level. The limits on
           $f_{M,2,3}/{\Lambda^4}$ can be determined using the relations: $f_{M,2} = 2 \times f_{M,0}$ 
           and $f_{M,3} = 2 \times f_{M,1}$.}
 \label{tab:aQGClimitsFM}
\end{table}
 \begin{figure}
  \centering
  \begin{tabular}{cc}
   \includegraphics[width=0.5\textwidth]{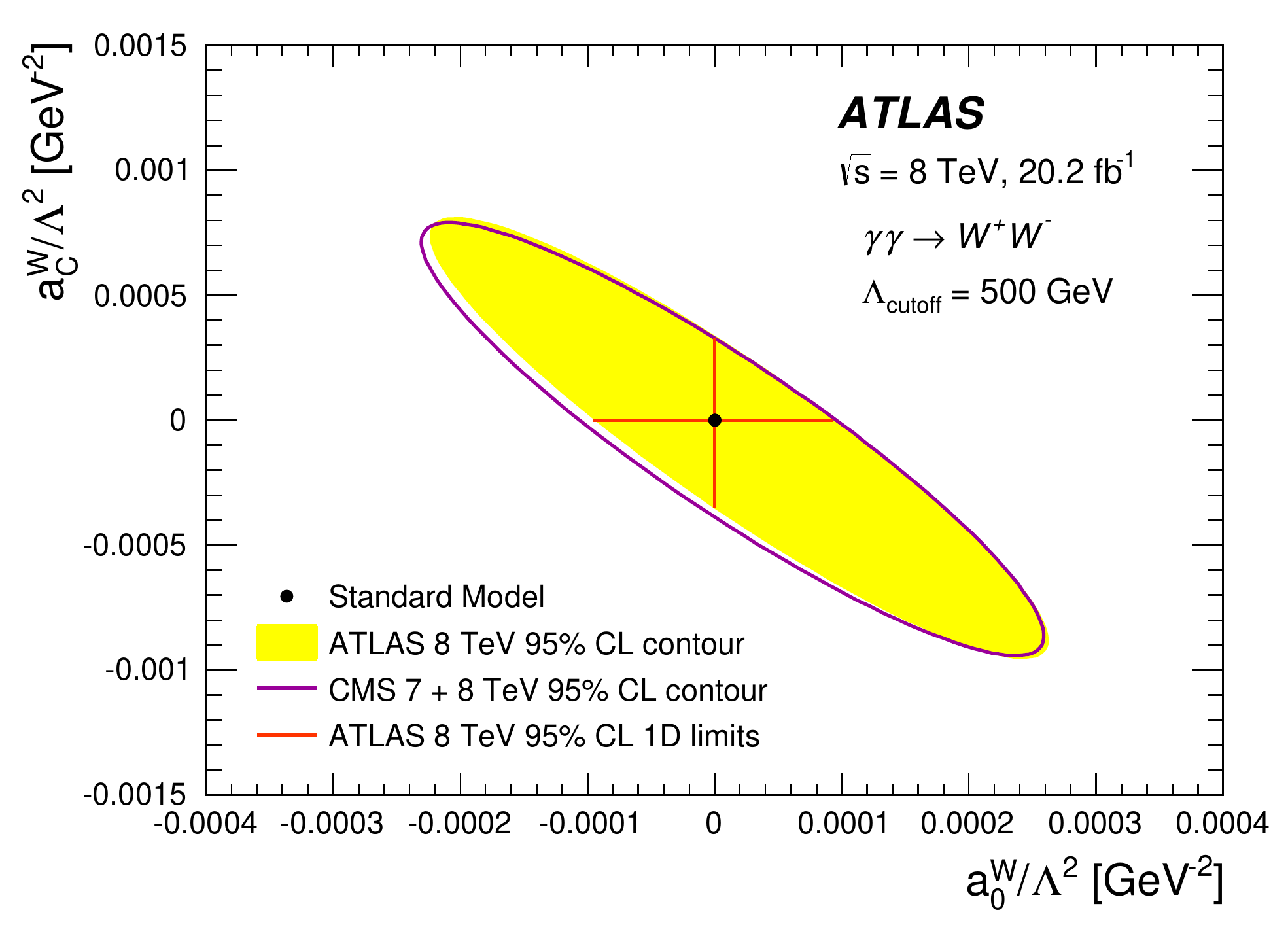} 
  \end{tabular}
  \caption{The observed log-likelihood 95\% confidence-level contour and 1D limits for the case with a dipole form factor 
           with $\Lambda_{\textrm{cutoff}} = 500~\GeV$. The CMS combined 7 and 8~\TeV result~\cite{CMSyyWW7and8} is shown for comparison.}
  \label{fig:aQGCcontour}
 \end{figure}
\par The 95\% CL limits on the dimension-8 $f_{M,0,1,2,3}/\Lambda^{4}$
couplings are given 
in Table~\ref{tab:aQGClimitsFM} for the cases with and without a form factor. 
They are derived from the $a_{0,C}^W/\Lambda^2$ couplings using Equation~(\ref{eqn:relationD8toD6}).
\subsection{Limits on exclusive Higgs Boson production}
\label{sec:HiggsResult}
\par As described in Section~\ref{sec:dataMCsamples},
exclusive production of Higgs bosons is simulated using
the FPMC generator. Exclusive $\WW$ contamination
 in the exclusive Higgs boson signal region is estimated by using \HERWIGPP 
samples that are scaled by $\fgam$ = 3.30 to account for
single-dissociative and double-dissociative processes.
The predicted background from exclusive $\WW$ is derived from the observed cross-section in the exclusive $\WW$ signal 
region (Section~\ref{sec:SMResults}).
As discussed in Sections~\ref{sec:MCTrackModelling}--\ref{sec:inclWWRegion}, the estimate for
inclusive $\WW$ and minor contributions of $\DrellYantotautau$ 
and $W+$jets is obtained from the inclusive \WW\ samples scaled by a
factor of 0.79. The contribution from inclusive Higgs production is
expected to be negligible. 
Exclusive dileptons are not scaled by $\fgam$ because LPAIR 
simulates SD and DD processes as discussed in Section~\ref{sec:validationDilepton}, 
except for \yytautau\ production of which only SD is simulated. 
The rest of the background sources are scaled by their respective 
correction factors to account for the mismodeling of the underlying event. 
Six candidates are observed in the data, while $3.0 \pm 0.8$ events are predicted from background,
and $0.023 \pm 0.003$ from signal. The quoted uncertainty is the sum in quadrature of systematic uncertainties.
Table~\ref{tab:Higgsyields} summarizes expected and observed yields in the signal region and at earlier 
selection points in the selection criteria summarized in Table~\ref{tab:selections}. The exclusive Higgs boson prediction 
quoted here is from elastic contribution only. Observed data reasonably agrees with predictions.
Figure~\ref{fig:mTSig} shows kinematic distributions in the signal region. 
\begin{table}[!h]
\centering
        \resizebox{\textwidth}{!}{
\providecommand{\cutflowTitle}{}
\begin{tabular}{l||rrr|rrr}
\hline\hline
 																							& Excl. $H$ Signal				& Data 		& Total Bkg 		& Incl. $\WW$ 	& Excl. $W^{+}W^{-}$ & Other Bkg 								 \\
\hline\hline
Preselection 																	& 0.065 $\pm$ 0.005 			& 129018 	& 120090				& 12844 				& 43  								& 107200\\
 $\pTemu>$30~\GeV, $\memu<55$~\GeV, $\dFem<1.8$ & 0.043 $\pm$ 0.004 			& 18568 	& 17060  				& 2026 					& 5.7									& 15030 \\
\DZ\ requirement 															& 0.023 $\pm$ 0.003 			& 8 			& 4.7 $\pm$ 1.3 & 1.4 $\pm$ 0.5 & 3.1 $\pm$ 1.3       & 0.2 $\pm$ 0.1  \\
$\mT<140$~\GeV [Signal Region] 							& 0.023 $\pm$ 0.003  			& 6 			& 3.0 $\pm$ 0.8 & 1.0 $\pm$ 0.4 & 1.8 $\pm$ 0.8       & 0.2 $\pm$ 0.1  \\
\hline\hline
\end{tabular}
}
\caption{Summary of signal and background yields at different stages
  of the Higgs boson event selection. Only major background sources are listed explicitly. All 
the other background sources are summed up in the `Other'
category. For the background, the uncertainties are only shown for the yields after exclusivity selection, where they are 
relevant for the measurement. They include the systematic and
statistical components, added in quadrature.}
\label{tab:Higgsyields}
\end{table}
\begin{figure}
 \includegraphics[width=0.5\linewidth]{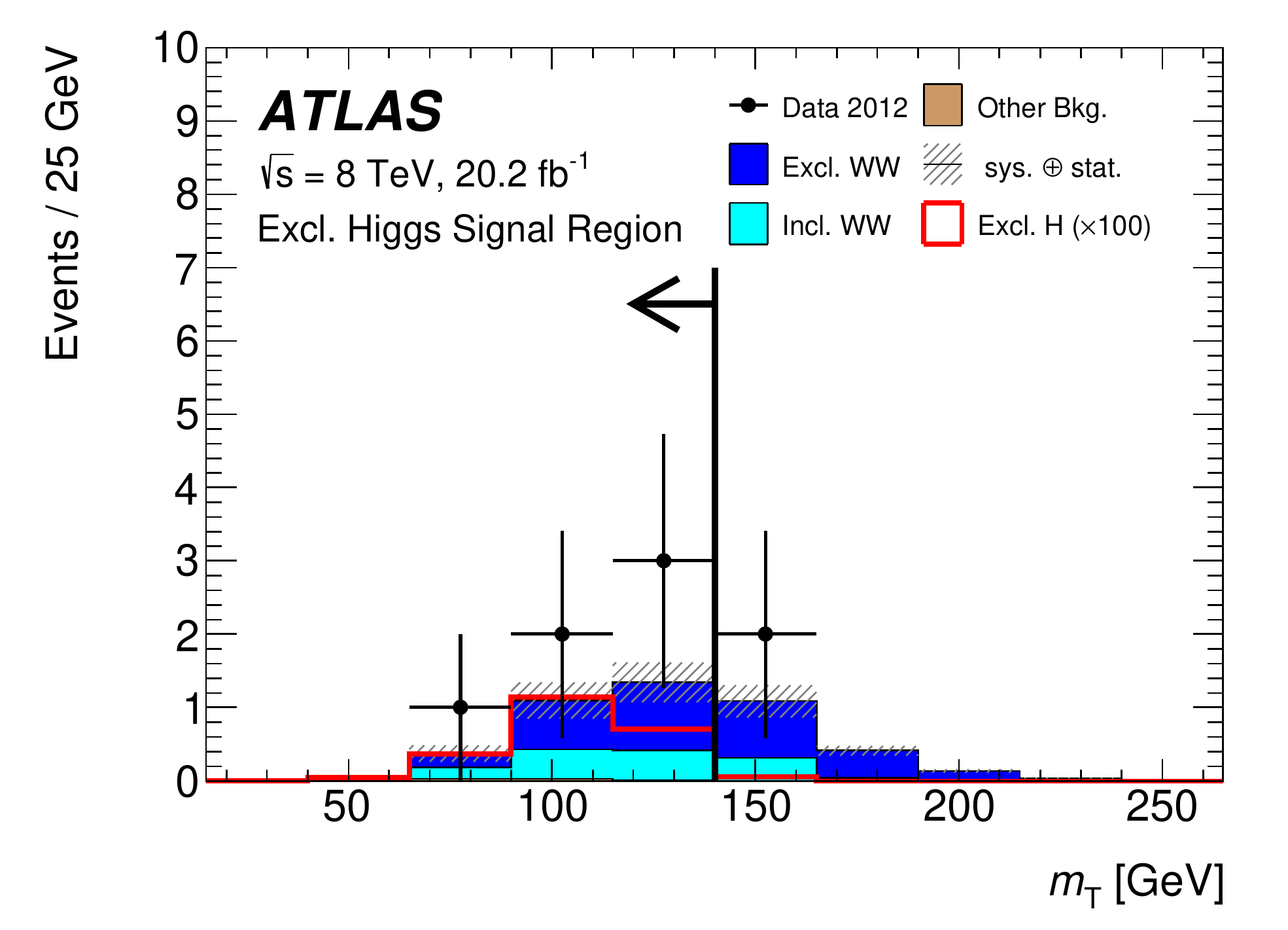} 
 \includegraphics[width=0.5\linewidth]{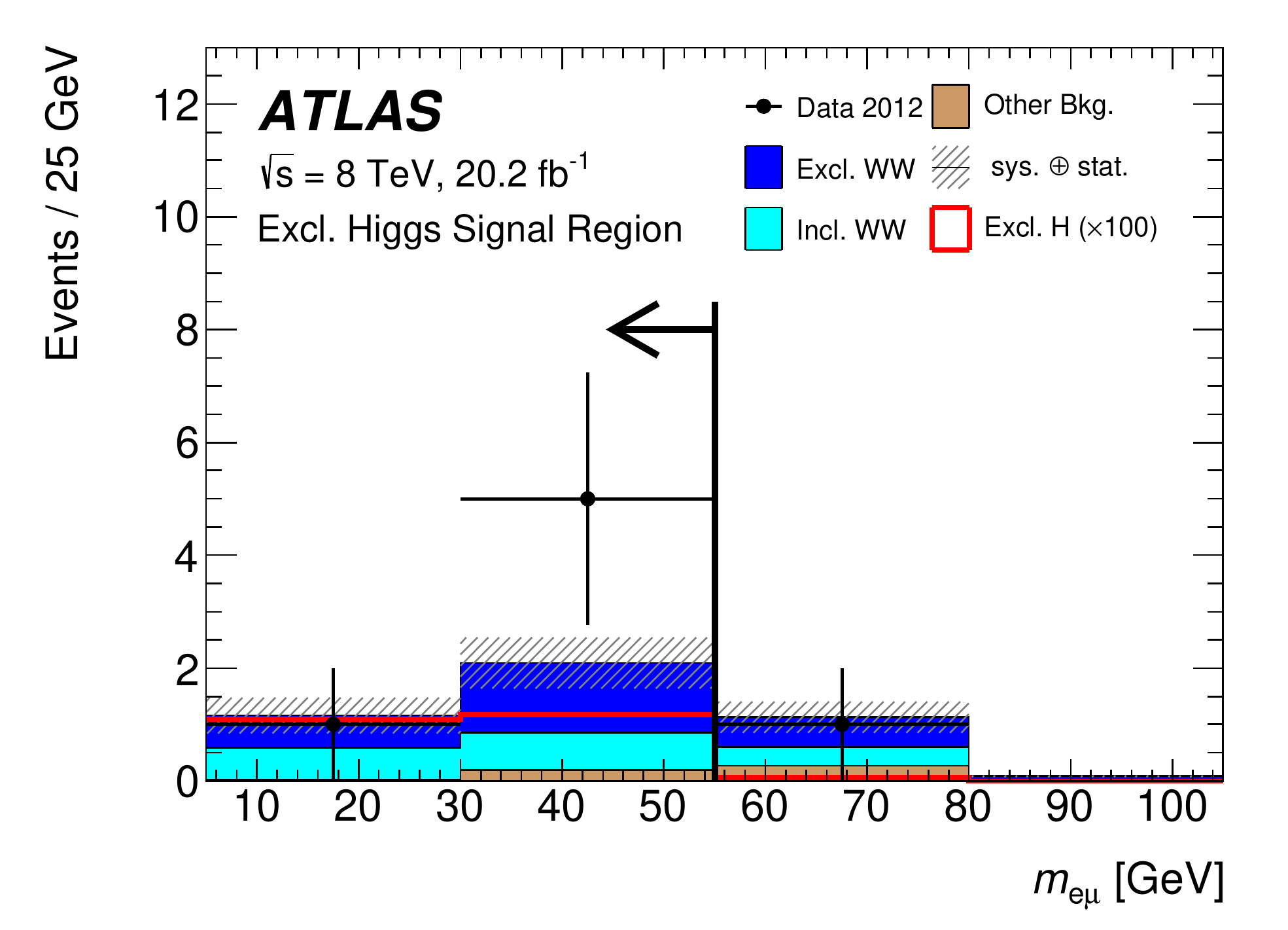}\\ 
\caption{Distributions in the exclusive Higgs boson signal region,
  without including the selection on the variable plotted. The dominant processes are inclusive and 
exclusive $\WW$ production. The expected signal is scaled by a factor
of 100 for visibility. The arrows denote the selection.}
\label{fig:mTSig}
\end{figure}
\par Yields summarized in the preceding paragraph are converted to
upper limits on the exclusive Higgs boson total production 
cross-section using the $CL_{S}$ technique~\cite{0954-3899-28-10-313}.
The branching ratio $BR(H\to\WW)$ used to compute these limits is 
$(21.5 \pm 0.9)$\%~\cite{Dittmaier:2012vm}.
Table~\ref{tab:models} shows a summary of the 95\% CL upper limits on the exclusive 
Higgs boson total production cross-section.
The observed upper limit is 1.2 pb, which is 1.1$\sigma$ higher than the expected 
upper limit of 0.7 pb. 
The statistical uncertainty in the predicted background dominates the
uncertainty involved in calculating this upper limit, 
while systematic uncertainties worsen the upper limits by at most 10\%.
This upper limit value is 400 times the cross-section predicted~\cite{Khoze}.
However, the limit would not change if the model prediction, which is for
elastic production only, increased by an order of magnitude. This limit calculation 
inherently assumes that the acceptance and efficiency for dissociative
events is not significantly different than for elastic events, hence
the associated systematic uncertainty is insignificant.
\begin{table}
\centering
\begin{tabular}{|cc|c|cc||r|}
\hline
$+2\sigma$ [pb] & $+1\sigma$ [pb] & Expected [pb] & $-1\sigma$ [pb] & $-2\sigma$ [pb] & Observed [pb] \\
\hline\hline
 1.6 & 1.0  & 0.7 &  0.5 &  0.4 & 1.2 \\
\hline 
\end{tabular}
\caption{Upper limits on $\sigma_H$ [pb] at 95\% CL. The $\pm 1\sigma$ and $\pm 2\sigma$ uncertainties quoted here are on the expected upper limit.}
\label{tab:models}
\end{table}

\section{Conclusion}
\label{sec:conclusion}
\par A measurement of the exclusive $\WW$ production cross-section 
and a search for exclusive Higgs boson production via diffraction
using $e^{\pm}\mu^{\mp}$
final states are presented using a data sample that corresponds to 20.2~\ifb\ 
of LHC $pp$ collisions at $\sqrt{s}=8$~\TeV collected with the ATLAS
detector. A track-based technique for selecting exclusive 
candidates was developed and validated in the $\mu^{+}\mu^{-}$ final  
state, resulting in a ratio of data to the EPA prediction for the exclusive
\yyll\ process of $\fEL = 0.76\pm 0.04 (\mathrm{stat.}) \pm 0.10 (\mathrm{sys.})$ 
in agreement with previous ATLAS measurements
at $\sqrt{s}=7~\TeV$. For exclusive $\WW$ production, the cross-section is determined to be 
$\sigma (\gamma\gamma\rightarrow W^{+}W^{-}\rightarrow e^{\pm}\mu^{\mp} X)
= 6.9 \pm 2.2 (\text{stat.}) \pm 1.4 (\text{sys.})$~fb
from 23 observed candidates with $8.3\pm2.6$ predicted background events. While evidence 
of SM exclusive $\WW$ production is at the $3.0\sigma$ level, no
evidence for an excess was seen in the 
kinematic region that would be enhanced by anomalous quartic gauge
couplings.  Rather, 
independent limits are placed on anomalous quartic gauge couplings
that are more stringent than earlier published results from the OPAL,
D0, and CMS experiments. Six
candidates consistent with exclusive 
Higgs boson production are observed in the data, with an expected SM
background of $3.0 \pm 0.8$ events. This result 
corresponds to an upper limit at 95\% CL on the total production cross-section 
of the exclusive Higgs boson of 1.2~pb, whereas the expected limit is 0.7~pb.  

\section*{Acknowledgments}
We thank CERN for the very successful operation of the LHC, as well as the
support staff from our institutions without whom ATLAS could not be
operated efficiently.
We acknowledge the support of ANPCyT, Argentina; YerPhI, Armenia; ARC, Australia; BMWFW and FWF, Austria; ANAS, Azerbaijan; SSTC, Belarus; CNPq and FAPESP, Brazil; NSERC, NRC and CFI, Canada; CERN; CONICYT, Chile; CAS, MOST and NSFC, China; COLCIENCIAS, Colombia; MSMT CR, MPO CR and VSC CR, Czech Republic; DNRF and DNSRC, Denmark; IN2P3-CNRS, CEA-DSM/IRFU, France; GNSF, Georgia; BMBF, HGF, and MPG, Germany; GSRT, Greece; RGC, Hong Kong SAR, China; ISF, I-CORE and Benoziyo Center, Israel; INFN, Italy; MEXT and JSPS, Japan; CNRST, Morocco; FOM and NWO, Netherlands; RCN, Norway; MNiSW and NCN, Poland; FCT, Portugal; MNE/IFA, Romania; MES of Russia and NRC KI, Russian Federation; JINR; MESTD, Serbia; MSSR, Slovakia; ARRS and MIZ\v{S}, Slovenia; DST/NRF, South Africa; MINECO, Spain; SRC and Wallenberg Foundation, Sweden; SERI, SNSF and Cantons of Bern and Geneva, Switzerland; MOST, Taiwan; TAEK, Turkey; STFC, United Kingdom; DOE and NSF, United States of America. In addition, individual groups and members have received support from BCKDF, the Canada Council, CANARIE, CRC, Compute Canada, FQRNT, and the Ontario Innovation Trust, Canada; EPLANET, ERC, FP7, Horizon 2020 and Marie Sk{\l}odowska-Curie Actions, European Union; Investissements d'Avenir Labex and Idex, ANR, R{\'e}gion Auvergne and Fondation Partager le Savoir, France; DFG and AvH Foundation, Germany; Herakleitos, Thales and Aristeia programmes co-financed by EU-ESF and the Greek NSRF; BSF, GIF and Minerva, Israel; BRF, Norway; Generalitat de Catalunya, Generalitat Valenciana, Spain; the Royal Society and Leverhulme Trust, United Kingdom.
The crucial computing support from all WLCG partners is acknowledged gratefully, in particular from CERN, the ATLAS Tier-1 facilities at TRIUMF (Canada), NDGF (Denmark, Norway, Sweden), CC-IN2P3 (France), KIT/GridKA (Germany), INFN-CNAF (Italy), NL-T1 (Netherlands), PIC (Spain), ASGC (Taiwan), RAL (UK) and BNL (USA), the Tier-2 facilities worldwide and large non-WLCG resource providers. Major contributors of computing resources are listed in Ref.~\cite{ATL-GEN-PUB-2016-002}.


\printbibliography


\clearpage

\input{atlas_authlist}

\end{document}

%% file: atlas_authlist.tex
\begin{flushleft}
{\Large The ATLAS Collaboration}

\bigskip

M.~Aaboud$^\textrm{\scriptsize 135d}$,
G.~Aad$^\textrm{\scriptsize 86}$,
B.~Abbott$^\textrm{\scriptsize 113}$,
J.~Abdallah$^\textrm{\scriptsize 64}$,
O.~Abdinov$^\textrm{\scriptsize 12}$,
B.~Abeloos$^\textrm{\scriptsize 117}$,
R.~Aben$^\textrm{\scriptsize 107}$,
O.S.~AbouZeid$^\textrm{\scriptsize 137}$,
N.L.~Abraham$^\textrm{\scriptsize 149}$,
H.~Abramowicz$^\textrm{\scriptsize 153}$
% ...
\bigskip
\\
% ...
$^{12}$ Institute of Physics, Azerbaijan Academy of Sciences, Baku, Azerbaijan\\
% ...
$^{64}$ University of Iowa, Iowa City IA, United States of America\\
% ...
$^{86}$ CPPM, Aix-Marseille Universit{\'e} and CNRS/IN2P3, Marseille, France\\
% ...
$^{107}$ Nikhef National Institute for Subatomic Physics and University of Amsterdam, Amsterdam, Netherlands\\
% ...
$^{113}$ Homer L. Dodge Department of Physics and Astronomy, University of Oklahoma, Norman OK, United States of America\\
% ...
$^{117}$ LAL, Univ. Paris-Sud, CNRS/IN2P3, Universit{\'e} Paris-Saclay, Orsay, France\\
% ...
$^{135}$ $^{(a)}$ Facult{\'e} des Sciences Ain Chock, R{\'e}seau Universitaire de Physique des Hautes Energies - Universit{\'e} Hassan II, Casablanca; $^{(b)}$ Centre National de l'Energie des Sciences Techniques Nucleaires, Rabat; $^{(c)}$ Facult{\'e} des Sciences Semlalia, Universit{\'e} Cadi Ayyad, LPHEA-Marrakech; $^{(d)}$ Facult{\'e} des Sciences, Universit{\'e} Mohamed Premier and LPTPM, Oujda; $^{(e)}$ Facult{\'e} des sciences, Universit{\'e} Mohammed V, Rabat, Morocco\\
% ...
$^{137}$ Santa Cruz Institute for Particle Physics, University of California Santa Cruz, Santa Cruz CA, United States of America\\
% ...
$^{149}$ Department of Physics and Astronomy, University of Sussex, Brighton, United Kingdom\\
% ...
$^{153}$ Raymond and Beverly Sackler School of Physics and Astronomy, Tel Aviv University, Tel Aviv, Israel\\
% ...
$^{a}$ Also at Department of Physics, King's College London, London, United Kingdom\\
$^{b}$ Also at Institute of Physics, Azerbaijan Academy of Sciences, Baku, Azerbaijan\\
$^{c}$ Also at Novosibirsk State University, Novosibirsk, Russia\\
$^{d}$ Also at TRIUMF, Vancouver BC, Canada\\
$^{e}$ Also at Department of Physics {\&} Astronomy, University of Louisville, Louisville, KY, United States of America\\
% ...
\end{flushleft}
